\documentclass[aps,prd,10pt,nofootinbib,floatfix]{revtex4-2}
\usepackage[titletoc]{appendix}
\usepackage[english]{babel}
\usepackage{caption}
\usepackage{float}
\usepackage{amssymb}
\usepackage{centernot}
\usepackage{upgreek}
\usepackage{yfonts}
\usepackage[fleqn]{amsmath}
\usepackage[T1]{fontenc}
\usepackage[x11names]{xcolor}
\usepackage{empheq}
\usepackage{color}
\allowdisplaybreaks
\usepackage{mathtools}
\usepackage{eucal}
\usepackage{graphicx}
\usepackage{graphics}
\usepackage{epstopdf}
\usepackage[section]{placeins}
\usepackage{bookmark}
\usepackage{xurl}
\usepackage{epsfig}
\usepackage{bm}
\usepackage[utf8]{inputenc} 
\usepackage{emerald}
\usepackage{slantsc}
\usepackage{array}

\makeatletter
\newcommand{\mathleft}{\@fleqntrue\@mathmargin0pt}
\newcommand{\mathcenter}{\@fleqnfalse}
\newsavebox\myboxA
\newsavebox\myboxB
\newlength\mylenA
\newcommand*\xoverline[2][0.75]{%
    \sbox{\myboxA}{$\m@th#2$}%
    \setbox\myboxB\null
    \ht\myboxB=\ht\myboxA%
    \dp\myboxB=\dp\myboxA%
    \wd\myboxB=#1\wd\myboxA
    \sbox\myboxB{$\m@th\overline{\copy\myboxB}$}
    \setlength\mylenA{\the\wd\myboxA}
    \addtolength\mylenA{-\the\wd\myboxB}%
    \ifdim\wd\myboxB<\wd\myboxA%
       \rlap{\hskip 0.5\mylenA\usebox\myboxB}{\usebox\myboxA}%
    \else
        \hskip -0.5\mylenA\rlap{\usebox\myboxA}{\hskip 0.5\mylenA\usebox\myboxB}%
    \fi}
\makeatother

\newcommand{\NOPRINT}[1]{\null}
\newcommand{\w}[1]{\hphantom{#1}}

\def\sst{\scriptscriptstyle}
\def\be{\begin{equation}}
\def\ee{\end{equation}}
\def\ba{\begin{eqnarray}}
\def\bal{\begin{eqnarray}\label}
\def\ea{\end{eqnarray}}
\def\bse{\begin{subequations}}
\def\ese{\end{subequations}}

\def\A{{\sst{\rm A}}}
\def\B{{\sst{\rm B}}}
\def\C{{\sst{\rm C}}}
\def\E{{\sst{\rm E}}}
\def\G{{\sst{\rm G}}}
\def\L{{\sst{\rm L}}}
\def\S{{\sst{\rm S}}}

\def\pd{\partial}
\def\a{{\mathsf{a}}}
\def\e{{\mathsf{e}}}

\def\n{{\mathsf{n}}}

\begin{document}
\title{Lunar Time in General Relativity}
\author{Sergei M. Kopeikin}
\affiliation{Department of Physics \& Astronomy, University of Missouri, 322 Physics Bldg., Columbia, Missouri 65211, USA}
\email{E-mail: kopeikins@missouri.edu}
\author{George H. Kaplan}
\affiliation{Contractor, U.S. Naval Observatory, 3450 Massachusetts Ave NW, Washington, DC 20392, USA}
\email{E-mail: george.h.kaplan.ctr@us.navy.mil}
\date{\today}
\begin{abstract}
We introduce the general-relativistic definition of Lunar Coordinate Time (TCL) based on the IAU 2000 resolutions that provide a framework for relativistic reference systems. From this foundation, we derive a transformation equation that describes the relative rate of TCL with respect to Geocentric Coordinate Time (TCG) {for various locations of the clock on lunar surface}. This equation serves as the cornerstone for constructing a relativistic TCL--TCG time conversion algorithm. Using this algorithm, we can compute both secular and periodic variations in the rate of an atomic clock placed on the Moon, relative to an identical clock on Earth. The algorithm accounts for various effects, including time dilation caused by the Moon's orbital motion around Earth, gravitational potentials of both Earth and Moon, and direct and indirect time dilation effects due to tidal perturbations caused by the Sun {and other major planets of the solar system}. Our approach provides exquisite details of the TCL--TCG transformation, {achieving a precision of several nanoseconds within the spatial volume dominated by the Earth's gravitational field known as the Hill sphere. This sphere extends from the Earth to a distance of approximately $1.5\times 10^6$ km, substantially encompassing the Moon's orbit.} To validate our methodology for lunar coordinate time, we compare it with the {mathemtical} formalism of local inertial frames applied to the Earth-Moon system, and confirm their equivalence. 
\end{abstract}
\maketitle
\tableofcontents
\section{Introduction}\label{s0}                 

Within the next decade, it is likely that humans will return to the lunar surface.  Any extended operations on the lunar surface, including exploration, surveying, construction, mining, and scientific work will require that standards be established for communication, lunar geodesy, navigation, and timekeeping (see, e.g.,  \citep{lunanet2022}).  It is reasonable, therefore, to anticipate that precise clocks may eventually be placed on the lunar surface.   One idea is to synchronize such clocks with Coordinated Universal Time (UTC) on the Earth through a microwave radio link \citep{gibney2023}, using two-way time transfer techniques well established for synchronizing Earth-based clock systems.

Besides the technical challenges of the clock synchronization procedure, there is a fundamental difference between time measured on Earth and that measured on the Moon, as described by Einstein's theory of general relativity.  The Moon has a weaker gravitational potential than the Earth and moves around the common center of mass, the Earth-Moon barycenter.  The Earth also moves around Earth-Moon barycenter, but because mass of the Earth is 81.3 times the mass of the Moon, this motion is less important; the Earth-Moon barycenter is actually within the Earth.  These two factors --- the different strengths of the gravitational fields and the relative motion --- cause a clock on the Moon to run at a different average rate relative to an identical clock on the Earth, even though both clocks advance by local SI seconds.  Furthermore, general relativity predicts periodic variations in the relative rate of lunar time with respect to Earth time. The amplitude of the periodic variations is an important factor in the choice of a synchronization scheme and practical metrological realization of the lunar time scale. 

As Moon and Earth move around the Sun, clocks on the Moon and Earth experience both relative rate differences and rather large periodic differences with respect to the barycentric coordinate time (called TCB) of the solar system.  The major monthly periodicity has an amplitude of 127\,$\mu$s between the two clocks from the point of view of the observer at rest with respect to the solar system barycenter. {{Physically, this periodicity comes from the relativity of the simultaneity of two events that are spatially separated in the GCRS, and its amplitude corresponds to the term $\frac{1}{c^2}{\bm v}_\E\cdot{\bm r}_\E$ in the TCB--TCG transformation \eqref{1}, computed for a hypothetical clock at the center of mass of the Moon.}
However, we expect this difference to be significantly smaller for an observer on the Earth. The physical reason to expect such a decrease in the amplitude of the periodic variation is that the Earth and the Moon form a single, gravitationally-bound system, which falls freely in the gravitational field of the Sun. The principle of equivalence applied to the Earth-Moon system states that the gravitational field of the Sun (and other planets of the solar system) can reveal itself only in the form of tidal terms which exclude the orbital (absolute) motion of the Earth-Moon system.   What really matters for the
comparison of the rate of clocks on Moon and Earth is merely the relative strength of the gravitational potentials of the Earth and Moon and the speed of their relative orbital motion, while the gravitational field of the Sun appears only in the form of the tidal potential having significantly smaller effect. This qualitative representation of the gravitational physics of the Earth-Moon system is well understood \citep{xiekop_2010,kopeikin_2010CeMDA,Ashby_2024} but insufficiently developed in view of the high importance of the problem for near-future practical applications for lunar science and technology. In order to reach significant reduction of errors in the maintenance of the lunar time and navigation system, it is necessary to better understand and accurately calculate the relativistic effects affecting the relative rate of terrestrial and lunar clocks. 

{The objective of this paper is to investigate the specific problem of clocks placed on the surfaces of the Earth and the Moon. A closely related application would be the creation of a GPS-like navigation satellite system around the Moon. The proper time maintained by clocks onboard these satellites  would be referenced to the coordinate time on the Moon's surface through additional relativistic transformations. While the general form of this proper-to-coordinate time transformations is known \citep{kopeikin_2011book}, they are not discussed in this paper as they fall outside the primary focus of our study.  Similarly, we do not discuss here the time kept by clocks on spacecraft at the Earth-Moon Lagrange points.}

The International Astronomical Union (IAU) has recommended the fundamental concepts, terminology, and basic equations of a relativistic framework of astronomical reference systems and time scales in the solar system \citep{iau2000}.  These have been firmly established and are described in a number of papers and textbooks, for example, \citep{Kopejkin_1988CeMec,bk_1990,soffel2003,kopeikin_2011book,Soffel_2013book}. Therefore, in this paper, we skip the mathematical description of the spacetime manifold, the metric tensor, and the principles of the matched asymptotic expansion technique.  The latter technique is used for finding solutions of the Einstein equations and the derivation of the coordinate transformations between global and local coordinates associated with different massive bodies of the solar system. 

After introducing the basic notations and conventions in Section \ref{noconastr}, we begin this paper by describing two crucial constructs that are part of the IAU framework.  In Section \ref{s-bct}, we consider the Barycentric Celestial Reference System (BCRS) and Barycentric Coordinate Time (TCB) for the solar system, and in Section \ref{s1}, we describe the Geocentric Celestial Reference System (GCRS) and Geocentric Coordinate Time (TCG) for the near-Earth environment.  We then explore the relationship between TCG and TCB.  Moving on, in Section \ref{s2}, we define the Lunar Celestial Reference System (LCRS) and its time counterpart, Lunar Coordinate Time (TCL), such that the definition of TCL closely mirrors that of TCG. Section \ref{s4} employs Einstein's principle of equivalence to minimize the gravitational influence of external solar system bodies on the Earth-Moon system by reducing it to tidal effects. In the context of Section \ref{s4}, we explore the relationship between the proper times $\tau_\E$ and $\tau_\L$, as would be measured by actual (although idealized) atomic clocks on the surfaces Earth and the Moon, with the corresponding coordinate times TCG and TCL. Additionally, we present a physical model for comparing TCL and TCG.  Section \ref{s5} discusses the influence of tidal effects on the evolution of the orbital elements of the Earth-Moon system. This sets the stage for developing an analytic model of the {lunar-to-terrestrial time} transformation in Section \ref{s5a}. The model takes into account several factors: (1)~the kinematic time dilation due to the relative motion of the Earth and Moon; (2)~the gravitational time dilation caused by gravitational potentials of the Earth and Moon; and (3)~the tidal time dilation due to the gravitational field of the Sun.  Section \ref{s7} reports on a number of numerical experiments that provide time-difference series that illustrate the complexity of the difference between time kept by clocks on the Moon and those on the Earth, the various components of which can be compared to the analytic model.  Finally, in Section \ref{s9}, we summarize our results and discuss several implications of these results for establishing a practical time standard for operations on or near the Moon. In the appendix \ref{appendix} of our work, we present a complementary theoretical model for lunar coordinate time. {This model follows recent theoretical development of the lunar time proposed by \citet{Ashby_2024}. The model leverages intermediate local coordinates associated with the Earth-Moon barycenter (EMCRS).  By doing so, we can directly compare our concept of lunar coordinate time with the one introduced by \citet{Ashby_2024}.} Remarkably, we find out that the two concepts are identical. However, the formalism based on the IAU 2000 resolutions provides a more comprehensive framework, allowing us to account for a greater number of terms in the TCL--TCG transformation.  \\

\section{Notations, Conventions and Astronomical Constants}\label{noconastr}    
We denote $c$ {= 299 792 458 m$\cdot$ s$^{-1}$} for the fundamental speed of the Minkowski spacetime which is equal to the speed of both light and gravity in general relativity. The symbol $G$  {= 6.67430$\times 10^{-11}$ m$^3\cdot$kg$^{-1}\cdot$ s$^{-2}$} denotes the universal gravitational constant. We use capital Roman subscripts (and indices) A, B, C, \ldots, to represent all massive bodies of the solar system. Subscripts E, L, S respectively denote Earth, Moon (L stands for Luna which is Latin for Moon), and Sun. Symbol $M_\A$ represents the mass of body A and $\mu_\A:= GM_\A$ represents the gravitational parameter of body A.

Throughout the paper we employ vector notations such that bold letters denote three-dimensional Cartesian vectors, for example, ${\bm a}=\left(a^1,a^2,a^3\right)$, ${\bm b}=\left(b^1,b^2,b^3\right)$. For any two vectors, let say, ${\bm a}$ and ${\bm b}$:
\begin{itemize}
\item[--] ${\bm a}\cdot{\bm b}$ denotes the Euclidean dot product of two vectors ${\bm a}$ and ${\bm b}$,
\item[--] ${\bm a}\times{\bm b}$ denotes the Euclidean cross product of two vectors ${\bm a}$ and ${\bm b}$.
\end{itemize}
Notations used for time scales are:
\begin{itemize}
\item[--] $t$ is the barycentric coordinate time TCB,
\item[--] $u$ is the geocentric coordinate time TCG,
\item[--] $s$ is the lunar coordinate time TCL,
\item[--] $\tau_\E$ is the proper time of clock located on the Earth {surface at a position which will be specified later},
\item[--] $\tau_\L$ is the proper time of clock located on the Moon {surface at a position which will be specified later}\,.
\end{itemize}
Several coordinate systems are used in the paper. They are the Barycentric Celestial Reference System (BCRS), Geocentric Celestial Reference System (GCRS), and Lunar Celestial Reference System (LCRS).  The corresponding coordinates are denoted as follows:
\begin{itemize}
\item[--] ${\bm x}$ is a barycentric (BCRS) radius-vector from the center of mass of the solar system to a point with the BCRS coordinate ${\bm x}$,
\item[--] ${\bm w}$ is a geocentric (GCRS) radius-vector from the geocenter to a point with the BCRS coordinate ${\bm x}$, 
\item[--] ${\bm z}$ is a selenocentric (LCRS) radius-vector directed from Moon's center of mass to a point with the BCRS coordinate ${\bm x}$.
\end{itemize}
{All spatial coordinates we use are interconnected through the corresponding post-Newtonian spatial transformations, as detailed in sources such as \citep{soffel2003,kopeikin_2011book} and \citep{xiekop_2010}. However, these post-Newtonian transformations are not necessary for the objectives of this paper, as they are relevant only in the second-post-Newtonian approximation, which is beyond the scope of our discussion. Therefore, we will use spatial coordinates as in Newtonian theory, utilizing only the spatial Galilean transformations.
{Let ${\cal P}$ be a space-time event at which the time and space coordinates are compared. The event ${\cal P}$ has $(t,{\bm x})$ coordinates in the BCRS, $(u,{\bm w})$ in the GCRS, and $(s,{\bm z})$ in the LCRS, respectively. All subsequent post-Newtonian transformations of time scales and space coordinates are computed at the event ${\cal P}$. Specifically, we denote:}}
\begin{itemize}  
\item[--] ${\bm x}_\A$ is the radius vector of the center of mass of body A in BCRS coordinates,
\item[--] ${\bm r}_\A:={\bm x}-{\bm x}_\A$ is a radius-vector from body A to the point with BCRS coordinates ${\bm x}$ ,
\item[--] ${\bm r}_{\A\B}:={\bm x}_\A-{\bm x}_\B$ is the radius-vector directed from body B to body A.
\end{itemize}
These notations establish relationships between the coordinate systems:
$${\bm w}={\bm x}-{\bm x}_\E={\bm r}_\E\qquad ,\qquad {\bm z}={\bm x}-{\bm x}_\L={\bm r}_\L\;,$$
{which are taken between the two spatial points at the same instant $t$ of TCB time. These spatial Galilean translations are sufficient for our discussion of time scales.}

We use a dot above a function or a vector that depends on time to denote a total derivative with respect to time, e.g. $\dot{\bm x}=d{\bm x}/dt$.   Thus, 
\begin{itemize}
\item[--] ${\bm v}_\A:=\dot{\bm x}_\A$ is the velocity vector of the center of mass of body A in BCRS coordinates,
\item[--] ${\bm a}_\A:=\dot{\bm v}_\A$ is the acceleration vector of the center of mass of body A in BCRS coordinates,
\item[--] ${\bm v}_{\A\B}:={\bm v}_\A-{\bm v}_\B$ is the relative velocity vector of body A with respect to body B.
\end{itemize}  
We introduce special notations for vectors of the relative positions and velocities of the bodies in the Earth-Moon-Sun system to simplify some equations
\begin{itemize}
\item[--] ${\bm r}:={\bm r}_{\L\E}$ is the geocentric radius-vector of the Moon with respect to the Earth,
\item[--] ${\bm r}':={\bm r}_{\E\S}$ is the heliocentric radius-vector of the Earth with respect to the Sun,
\item[--] ${\bm v}:={\bm v}_{\L\E}$ is the geocentric velocity vector of the Moon, 
\item[--] ${\bm v}':={\bm v}_{\E\S}$ is the heliocentric velocity vector of the Earth.
\end{itemize}

For numerical estimates given in Section \ref{s5a} we use the following approximate values of the astronomical constants taken from NASA website \citep{nasafacts},
\begin{itemize}
\item mass of the Sun, $M_\S=1.9884\times 10^{30}$ kg
\item mass of the Earth, $M_\E=5.9722\times 10^{24}$ kg
\item mass of the Moon, $M_\L=7.346\times 10^{22}$ kg
\item the mean radius of the Earth, $R_\E=6.371\times 10^6$ m
\item the mean radius of the Moon, $R_\L=1.737\times 10^6$ m
\item semi-major axis of the lunar orbit, $a=3.8444\times 10^8$ m
\item semi-major axis of the Earth orbit, $a'=1.49598\times 10^{11}$ m
\item eccentricity of the lunar orbit, $e=0.0549$
\item eccentricity of Earth's orbit, $e'=0.0167$
\item inclination of lunar orbit to ecliptic, $i= 5.145^\circ=0.0898$ rad
\item inclination of lunar equator to ecliptic, $I=1.543^\circ=0.0269$ rad
\end{itemize}
Additional notations and conventions are introduced in the text of the paper and in Appendix \ref{appendix}.

\section{The Barycentric Coordinate Time}\label{s-bct}  

The concept of the relativistic time scales and reference coordinate systems was a matter of intensive theoretical discussion at various working groups, conferences and meetings of the IAU starting more than three decades ago (see, for example, \citep{bk_1990,sofbrum1991,guinot1995,capitaine1997}). The various technical issues were settled at meetings of several IAU working groups, where corresponding resolutions on time scales and reference systems had been formulated. The resolutions were finally adopted at the IAU General Assembly in 2000 \citep{iau2000} and are fully described
in \citep{soffel2003}. 

The primary astronomical time scale is Barycentric Coordinate Time $t:=$ TCB ({\it Temps-Coordonn\'ee Barycentrique}) of the solar system. It is a uniform time, {flowing at a constant rate, conceptually equivalent to the proper time measured by synchronized clocks of a set of fictitious observers at rest with respect to the solar system barycenter, with the gravitational field of all bodies of the solar system completely removed.  {\it Syst\'{e}me International} (SI) second is the basic unit \citep{kopeikin_2011book,Soffel_2013book}. In the language of general relativity this configuration is known as asymptotically-flat Minkowskian spacetime \citep{mtw}.} 

TCB is defined as the independent argument of the relativistic equations of motion of solar system bodies referred to the Barycentric Celestial Reference System (BCRS).  The solutions of these equations for the coordinates and velocities of the bodies are given either analytically \citep{Kudryavtsev-2016a,Kudryavtsev-2016b} or tabulated numerically  \citep{INPOP_2020,DE440_2021AJ} in the form of the solar system ephemerides \citep{Fienga_2024LRR}.  To the extent that the effects of the gravitational field of the Milky Way and other galaxies can be considered negligibly small and not sensible in the dynamics of the solar system bodies, the BCRS can be considered as a global inertial coordinate system. 

We denote BCRS spatial coordinates as ${\bm x}=\left(x^i\right)$ where $x^1:= x,\; x^2:= y,\; x^3:= z$. The origin of the BCRS is at the center of mass (barycenter) of the solar system bodies.  The spatial axes of BCRS are not rotating kinematically with respect to the set of distant quasars that form the International Celestial Reference Frame (ICRF) in the sky \citep{icrf2020}, which matches the Gaia catalog orientation with an uncertainty below 20~microarcseconds \citep{Malkin_2024}. 

\section{The Geocentric Coordinate Time}\label{s1}   

Geocentric Coordinate Time $u:=$ TCG ({\it Temps-Coordonn{\'e}e G{\'e}ocentrique}) is the coordinate time of the local Geocentric Celestial Reference System (GCRS). It can be interpreted as the proper time {measured by synchronized clocks of a set of fictitious observers at rest with respect to the geocenter, assuming the gravitational field of the Earth is removed} \citep{kopeikin_2011book}.  The geocenter is the center of mass of the Earth, including the oceans and atmosphere.  TCG is the uniform argument of the equations for the evolution of the Earth's rotational parameters (precession, nutation, polar wobble and diurnal rotation) as well as the equations of motion of artificial satellites in orbit around the Earth.

According to the IAU 2000 resolutions, Geocentric Coordinate Time (TCG) is related to the Barycentric Coordinate Time (TCB) of the solar system {through the following pointwise} relativistic equation \citep{iau2000,kopeikin_2011book}
\bal{1} 
u&=&t-\frac{1}{c^2}\int\limits_{t_0}^t\left(\frac{v_\E^2}{2}+\frac{{\mu}_\L}{r_{\E\L}}+\sum_{\A\neq\E,\L}\frac{{\mu}_\A}{r_{\E\A}}\right)dt-\frac{1}{c^2}{\bm v}_\E\cdot{\bm r}_\E\\\nonumber
&&{\;\;-\frac{1}{c^4}\int\limits_{t_0}^t\left(\frac{v_\E^4}{8}+\frac{3v_\E^2}{2}\sum_{\A\neq\E}\frac{{\mu}_\A}{r_{\E\A}}-4\sum_{\A\neq\E}\frac{{\mu}_\A}{r_{\E\A}}{\bm v}_\E\cdot{\bm v}_\A-\sum_{\A\neq\E}\sum_{\B\neq\E}\frac{{\mu}_\A\mu_\B}{r_{\E\A}r_{\E\B}}\right)dt+{\cal O}(c^{-5})\;,}
\ea
where $u$ denotes TCG, $t$ denotes TCB, sub-index L refers to the Moon, sub-index E refers to the Earth, and sub-index A refers to all other massive bodies of the solar system. In Eq. \eqref{1}, the point { (also referred to as an event in relativistic terms)} in space at which the time $u$ is calculated has the BCRS spatial coordinates ${\bm x}=\left(x^i\right)=\left(x^1,x^2,x^3\right)$, {and the TCB time $t$, which is also the instant of time at which the spatial coordinates of all bodies are determined}, so that ${\bm x}_\L:= {\bm x}_\L(t)$, ${\bm x}_\E:= {\bm x}_\E(t)$ and ${\bm x}_\A:= {\bm x}_\A(t)$. The BCRS velocities of the Earth's center of mass and all other bodies of the solar system are also computed at time $t$, that is ${\bm v}_\E:={\bm v}_\E(t)=d{\bm x}_\E/dt$, and ${\bm v}_\A:={\bm v}_\A(t)=d{\bm x}_\A/dt$. The BCRS coordinate distance between the Earth and a body A is denoted as $r_{\E\A}=|{\bm r}_{\E\A}|$, ${\bm r}_{\E\A}={\bm x}_\E-{\bm x}_\A$. The BCRS vector ${\bm r}_\E={\bm x}-{\bm x}_\E$ is directed from the Earth to the point where the time $u$ is calculated.

Notice that in accordance with the IAU resolutions, the overall gravitational potential within the integrals in Eq. \eqref{1} does not include the gravitational potential of the Earth. This omission arises from the general relativistic definition of the GCRS \citep{soffel2003}. According to this definition, the GCRS is considered to be in a state of free fall within the gravitational field of the external massive bodies of the solar system, which includes the Moon. However, it is instructive to single out the gravitational potential of the Moon $\mu_\L/r_{\E\L}=GM_\L/r_{\E\L}$ in Eq. \eqref{1} from the sum of the gravitational potentials over the other solar system bodies. This is because the Moon is gravitationally bound to Earth, forming the Earth-Moon system that itself is in the state of a free fall in the field of the other massive bodies of the solar system like Sun, Jupiter, etc. Unlike the external bodies, the Earth and the Moon experience a unique gravitational interplay. This property of the Earth-Moon system significantly amplifies the impact of the Moon's gravitational field on lunar time, setting it apart from the effects of the other external bodies. A more detailed exploration {and clarification of this statement} is given in the next section.

Eq. \eqref{1} is formulated within the framework of the {second post-Newtonian approximation (2~PNA), which incorporates terms of the order ${\cal O}(c^{-4})$. Only the main post-post-Newtonian terms have been included in \eqref{1}. Many additional post-post-Newtonian terms exist \citep{kopeikin_2011book,k2019PRD}, but they have been omitted as they are negligibly small for the problem under discussion and can be ignored for near-future practical applications. The explicit form of the residual 2~PNA terms in Eq. \eqref{1} can be found in publications \citep{bk_1990,Klioner_1993,kopeikin_2011book}. }

It is important to note that the TCG--TCB time transformation, as per Eq. \eqref{1}, is formulated under the approximation of spherically-symmetric and non-rotating bodies, known as the mass-monopole gravitational field approximation \citep{soffel2003}. The effects associated with the non-sphericity and rotation of the bodies can be incorporated into the time transformation \eqref{1} (refer to the book by \citet[sections 4, 5]{kopeikin_2011book} for more details). For example, when the Sun's quadrupole perturbation is included in the TCG--TCB transformation, as outlined by \citet{DE440_2021AJ}, it results in an additional secular divergence between the TCG and TCB time scales. Our calculations estimate this divergence to be less than $10^{-5}$ ns/day. Similarly, incorporating the quadrupole perturbation of the Moon's gravitational field results in a secular drift rate of TCG--TCB of less than $10^{-6}$ ns/day. The magnitude of these higher multipole perturbations of the TCG--TCB transformation \eqref{1} is too small to be important for timekeeping metrology and navigation in cislunar space. Therefore, the effects of the higher multipoles of the Moon, Sun, and other planets are not discussed further in this paper.

We denote GCRS spatial coordinates as ${\bm w}=\left(w^i\right)=\left(w^1,w^2,w^3\right)$. The origin of the GCRS is at the geocenter and the spatial axes of the GCRS are kinematically non-rotating with respect to the spatial axes of BCRS. This convention on the direction of the coordinate axes of the BCRS and GCRS introduces a small relativistic term to the GCRS metric tensor, caused by the geodetic precession of Earth's intrinsic angular momentum vector around the vector of Earth's orbital angular momentum \citep{kopeikin_2011book}. This remark is important in building the precise theory of the relativistic equations of motion but not essential for the derivation of the time transformations that appear in the next sections. 
  
\section{The Lunar Coordinate Time}\label{s2}     

The concept of local relativistic time scales associated with the other massive bodies of the solar system (planetocentric coordinate time) is similar in nature to that of TCG. It was introduced and briefly discussed in the work by \citet{bk_1990}. In particular, we can introduce the lunar coordinate time $s:=$ TCL ({\it Temps-Coordonn{\'e}e Lunocentrique}). It is related to TCB by a transformation which is similar to \eqref{1} where all `Earth' terms must be replaced with `lunar' terms and vice versa \citep{bk_1990,xiekop_2010}  
\bal{2} 
s&=&t-\frac{1}{c^2}\int\limits_{t_0}^t\left(\frac{v_\L^2}{2}+\frac{{\mu}_\E}{r_{\L\E}}+\sum_{\A\neq\E,\L}\frac{{\mu}_\A}{r_{\L\A}}\right)dt-\frac{1}{c^2}{\bm v}_\L\cdot{\bm r}_\L\\\nonumber
&&{\;\;-\frac{1}{c^4}\int\limits_{t_0}^t\left(\frac{v_\L^4}{8}+\frac{3v_\L^2}{2}\sum_{\A\neq\L}\frac{{\mu}_\A}{r_{\L\A}}-4\sum_{\A\neq\L}\frac{{\mu}_\A}{r_{\L\A}}{\bm v}_\L\cdot{\bm v}_\A-\sum_{\A\neq\L}\sum_{\B\neq\L}\frac{{\mu}_\A\mu_\B}{r_{\L\A}r_{\L\B}}\right)dt+{\cal O}(c^{-5})\;.}
\ea
Notice that, according to this definition, $r_{\L\E}=r_{\E\L}$ is the distance between the centers of mass of the Earth and the Moon, the BCRS vector, ${\bm r}_\L={\bm x}-{\bm x}_\L$, is directed from the center of mass of the Moon to the point ${\bm x}$ where the time $s$ is calculated, and ${\bm v}_\L=\dot{\bm x}_\L=d{\bm x}_\L/dt$ is the BCRS velocity of the Moon.

The TCL--TCB transformation as given in Eq. \eqref{2} is formulated in the mass-monopole approximation similar to the TCG--TCB transformation given by Eq. \eqref{1}. The most significant contribution from the omitted higher-order multipoles could come only from the quadrupole moment $J_2$ of the Earth's gravitational field. For example, the secular drift that is caused by the $J_2$ perturbation on GPS satellite time reaches 0.6~ns/day, while the periodic variations have an amplitude of 0.02~ns 
\citep{Ashby_2003LRR}.  Detailed study of this effect reveals that its inclusion improves the clock solution precision, as referenced in \citep{Kouba_2019GPSS,Formichella_2021,Wang_2023}. However, the amplitude of the quadrupole-induced perturbation scales inversely proportional to the square of the radius of the clock's orbit. Given that the radius of the lunar orbit is 15 times that of GPS satellites, the effect of $J_2$ on the secular drift of the lunar clock decreases to
0.004~ns/day, which is completely insignificant for the operation of that clock. 

The Earth-Moon system is a binary system bounded by the mutual gravitational force of attraction. For this reason, it is convenient to single out in Eq. \eqref{2} those terms that belong to the Earth-Moon system from those that belong to the external massive bodies. This is achieved by shifting the origin of BCRS to geocenter which leads to transformation of the relative radius vector,
\bal{3}
{\bm r}_{\L\A}&=&{\bm x}_{\L}-{\bm r}_{\A}=({\bm x}_{\L}-{\bm x}_{\E})+({\bm x}_{\E}-{\bm r}_{\A})={\bm r}_{\L\E}+{\bm r}_{\E\A}\;,
\ea
where ${\bm r}_{\L\E}={\bm x}_\L-{\bm x}_\E$ is the radius vector of the Moon with respect to the Earth. Taking the transformation of the BCRS coordinates of the Moon, ${\bm x}_\L={\bm x}_\E+{\bm r}_{\L\E}$ and differentiating it with respect to TCB yields the transformation of the lunar velocity,
\bal{3a}
{\bm v}_\L&=&{\bm v}_\E+{\bm v}_{\L\E}\;,
\ea
where ${\bm v}_{\L\E}:={\bm v}_{\L}-{\bm v}_{\E}$ is the relative velocity of the Moon with respect to the Earth. 

{Transformations \eqref{3} and \eqref{3a} allow us to consider the orbital motion of the Moon with respect to the Earth, but not with respect to the barycenter of the solar system. These transformations are essentially the Galilean coordinate transformations from the BCRS to GCRS, which are sufficient at the level of the first post-Newtonian approximation. We will analyze this in detail in the following sections.}

{More precise second post-Newtonian terms in Eqs. \eqref{1} and \eqref{2} require the inclusion of post-Newtonian corrections to Eqs. \eqref{3} and \eqref{3a}, complicating the overall mathematical analysis of lunar time. However, we have a valid physical reason to ignore these complications at this stage of our theoretical study. The Einstein principle of equivalence applied to the Earth-Moon system indicates that the effects of the gravitational field of the Sun and other major bodies of the solar system on the gravitational physics of the Earth-Moon system are reduced merely to tidal effects \citep{kopeikin_2007PhRvL,kopeikin_2010CeMDA}. This holds true not only in the first post-Newtonian approximation of the order of ${\mathcal O}\left(c^{-2}\right)$, as we will demonstrate below, but also in the second post-Newtonian approximation of the order of ${\mathcal O}\left(c^{-4}\right)$ --- see \citep{xiekop_2010} for more detail.}

{The only significant contribution to the relative rate of the lunar coordinate time (TCL) with respect to TCG in the second post-Newtonian approximation arises from terms that depend exclusively on either the internal characteristics of the Earth-Moon system or the products of tidal terms with the square of the orbital velocity of the Earth around the Sun \citep{brumkop1989NC,xiekop_2010}. Consequently, the contribution of the second post-Newtonian terms in the TCL-TCG transformation is of the order of either $(v_{\L\E}/c)^4$, $(v_{\L\E}/c)^2 (\mu_\E/c^2 r_{\L\E})$ or $(v_\E/c)^2(\mu_\S/c^2 r_{\E\S})(r_{\L\E}/r_{\S\E})^2$. These terms induce a secular drift in lunar time, but they are all on the order of a few times $10^{-8}$ ns/day, rendering them practically unnoticeable. These arguments, based on the principle of equivalence, demonstrate that all second post-Newtonian order terms are exceedingly small and will be ignored in the remainder of this paper.
}

The transformation \eqref{3} is used to expand the terms within the sum in Eq. \eqref{2} in a Taylor series with respect to a small parameter, which is the ratio of the Earth-Moon distance to the distance to the external bodies. More specifically,
\bal{4}
\frac{{\mu}_\A}{r_{\L\A}}&=&\frac{{\mu}_\A}{|{\bm r}_{\E\A}+{\bm r}_{\L\E}|}=\frac{{\mu}_\A}{r_{\E\A}}-\frac{{\mu}_\A}{r^3_{\E\A}}\left({\bm r}_{\E\A}\cdot{\bm r}_{\L\E}\right)+\frac32\frac{{\mu}_A}{r_{\E\A}^5}\left[\left({\bm r}_{\E\A}\cdot{\bm r}_{\L\E}\right)^2-\frac13r_{\E\A}^2r_{\L\E}^2\right]+{\cal O}\left(\frac{r_{\L\E}^3}{r_{\E\A}^3}\right)\;,
\ea
where the residual terms are negligibly small and will be dropped out from the subsequent equations. The Taylor expansion \eqref{4} of the gravitational potential is also known as the expansion in tidal multipoles \citep{kopeikin_2011book,soffel2019book}.
It follows from \eqref{4} that
\bal{5}
\sum_{\A\neq\E,\L}\frac{{\mu}_\A}{r_{\L\A}}&=&\sum_{\A\neq\E,\L}\frac{{\mu}_\A}{r_{\E\A}}-\sum_{\A\neq\E,\L}\frac{{\mu}_\A}{r^3_{\E\A}}\left({\bm r}_{\E\A}\cdot{\bm r}_{\L\E}\right)+\frac32\sum_{\A\neq\E,\L}\frac{{\mu}_\A}{r^5_{\E\A}}\left[\left({\bm r}_{\E\A}\cdot{\bm r}_{\L\E}\right)^2-\frac13r_{\E\A}^2r_{\L\E}^2\right]\\\nonumber\\\nonumber
&=&-\frac{{\mu}_\L}{r_{\L\E}}+\sum_{\A\neq\E,\L}\frac{{\mu}_\A}{r_{\E\A}}-\sum_{\A\neq\E}\frac{{\mu}_\A}{r^3_{\E\A}}\left({\bm r}_{\E\A}\cdot{\bm r}_{\L\E}\right)+\frac32\sum_{\A\neq\E,\L}\frac{{\mu}_\A}{r^5_{\E\A}}\left[\left({\bm r}_{\E\A}\cdot{\bm r}_{\L\E}\right)^2-\frac13r_{\E\A}^2r_{\L\E}^2\right]\;.
\ea
Notice that, in the above equation, we have included the lunar term (the term with the subscript L) in the second sum in the right-hand side of the equation. This was done to allow us to single out the terms depending on the BCRS orbital acceleration of the Earth, which are given in the Newtonian and point-like body approximation by the following equation:
\bal{6}
{\bm a}_\E&=&-\sum_{\A\neq\E}\frac{{\mu}_\A}{r^3_{\E\A}}{\bm r}_{\E\A}\;,
\ea
where the sum on the right-hand side depends on all masses outside the Earth, which includes the Moon. {By isolating the acceleration term \eqref{6}, we can identify a combination of terms in Eq. \eqref{2} that forms a total time derivative, which can then be integrated.  }

Accounting for \eqref{5} and \eqref{6} and omitting the tidal terms from external major planets (they are, at least, a factor of $10^{-3}$ smaller than that from the Sun), we can bring the time transformation Eq. \eqref{2} to a simpler form
\bal{7} 
s&=&t-\frac{1}{c^2}\int\limits_{t_0}^t\left\{\frac{v_\E^2}{2}+\frac{v_{\L\E}^2}{2}+\frac{{\mu}_\E}{r_{\L\E}}-\frac{{\mu}_\L}{r_{\L\E}}+\sum_{\A\neq\E,\L}\frac{{\mu}_\A}{r_{\E\A}}+\frac32\frac{{\mu}_\S}{r^5_{\E\S}}\left[\left({\bm r}_{\E\S}\cdot{\bm r}_{\L\E}\right)^2-\frac13r_{\E\S}^2r_{\L\E}^2\right]\right\}dt\\\nonumber
&&\;\;-\frac{1}{c^2}\int\limits_{t_0}^t\left({\bm v}_{\L\E}\cdot{\bm v}_\E+{\bm r}_{\L\E}\cdot{\bm a}_\E\right)dt-\frac{1}{c^2}{\bm v}_\L\cdot{\bm r}_\L
\;,
\ea
which makes the terms depending on the BCRS velocity and acceleration of the Earth integrable. Performing the integration of these terms in the second line of Eq. \eqref{7}, we get
\bal{8}
s&=&t-\frac{1}{c^2}\int\limits_{t_0}^t\left\{\frac{v_\E^2}{2}+\frac{v_{\L\E}^2}{2}+\frac{{\mu}_\E}{r_{\L\E}}-\frac{{\mu}_\L}{r_{\L\E}}+\sum_{\A\neq\E,\L}\frac{{\mu}_\A}{r_{\E\A}}+\frac32\frac{{\mu}_\S}{r^5_{\E\S}}\left[\left({\bm r}_{\E\S}\cdot{\bm r}_{\L\E}\right)^2-\frac13r_{\E\S}^2r_{\L\E}^2\right]\right\}dt
\\\nonumber
&&\;\;-\frac{1}{c^2}\Bigl({\bm v}_\E\cdot{\bm r}_{\L\E}+{\bm v}_\L\cdot{\bm r}_\L\Bigr)\;,
\ea
where the subscript S stands for the Sun. The terms with the BCRS coordinates and velocity of the Moon can be rearranged as follows
\bal{9}
{\bm v}_\E\cdot{\bm r}_{\L\E}+{\bm v}_\L\cdot{\bm r}_\L&=&{\bm v}_\E\cdot{\bm r}_\E+{\bm v}_{\L\E}\cdot{\bm r}_\E-{\bm v}_{\L\E}\cdot{\bm r}_{\L\E}\;.
\ea
so that the equation of transformation between lunar and barycentric times finally reads
\bal{10}
s&=&t-\frac{1}{c^2}\int\limits_{t_0}^t\left\{\frac{v_\E^2}{2}+\frac{v_{\L\E}^2}{2}+\frac{{\mu}_\E}{r_{\L\E}}-\frac{{\mu}_\L}{r_{\L\E}}+\sum_{\A\neq\E,\L}\frac{{\mu}_\A}{r_{\E\A}}+\frac32\frac{{\mu}_\S}{r^5_{\E\S}}\left[\left({\bm r}_{\E\S}\cdot{\bm r}_{\L\E}\right)^2-\frac13r_{\E\S}^2r_{\L\E}^2\right]\right\}dt
\\\nonumber
&&\;\;-\frac{1}{c^2}\Bigl({\bm v}_\E\cdot{\bm r}_\E+{\bm v}_{\L\E}\cdot{\bm r}_\E-{\bm v}_{\L\E}\cdot{\bm r}_{\L\E}\Bigr)\;.        
\ea
This form of the time transformation equation is the most convenient for comparing the ticking rate of TCL with respect to TCB. 

In what follows, we shall also use the Lunar Celestial Reference System (LCRS). Its spatial coordinates are denoted ${\bm z}=\left(z^i\right)=\left(z^1,z^2,z^3\right)$. The origin of LCRS is at the center of mass of the Moon and the spatial axes of LCRS are non-rotating with respect to the spatial axes of the BCRS.  It is the lunar analog of the GCRS.\footnote{Definitions for the LCRS and its coordinate time TCL, consistent with the usage
in this paper, were the subjects of a resolution passed by the XXXII IAU General Assembly in 2024 \citep{iau2024}.}

\section{Physical Model of the Time Comparison of Atomic Clocks on Earth and Moon}\label{s4}                

We can use the equations from the previous sections to compare TCG and TCL. The time transformations TCG--TCB \eqref{1} and TCL--TCB \eqref{10} are performed at the point at which the BCRS coordinates were denoted as ${\bm x}$. This is the point where the real clock will be placed. In practice, two  different clocks measuring time will be used --- one clock will be placed on the lunar surface and another clock is placed on the Earth. Hence, we have to distinguish the coordinates of the spatial locations of clocks in the TCL--TCB and TCG--TCB transformations. 

Let us label the clock counting the proper time on the Earth as the E-clock (Earth clock), and the clock counting the proper time on the Moon as the
L-clock (lunar clock).  The spatial position of the E-clock in the GCRS is ${\bm w}$ and the spatial position of the L-clock in the LCRS is ${\bm z}$. Notice that the E-clock counts its proper time denoted as $\tau_\E$, which is related to TCG by an additional transformation, and the L-clock counts its proper time denoted as $\tau_\L$, which is related to TCL by another additional transformation. These additional transformations from the proper time of each clock to a corresponding coordinate time will be given below in this section. 

In order to write down the TCL--TCG time transformation, we subtract the TCG time Eq. \eqref{1} from the TCL time Eq. \eqref{10}. It is straightforward to observe that most of the BCRS coordinate-related terms cancel out. This is because the principle of equivalence eliminates the main BCRS-dependent terms, leaving only the terms having a direct relation to the Earth-Moon system, while the contribution of the external bodies (mainly the Sun) appears in the form of tidal terms \citep{kopeikin_2011book,kopeikin_2010CeMDA}. Thus, the TCL--TCG difference  has a relatively simple expression,
\bal{11}
s&=&u-\frac{1}{c^2}\int\limits_{t_0}^t\left\{\frac{v_{\L\E}^2}{2}+\frac{{\mu}_\E}{r_{\L\E}}-\frac{2{\mu}_\L}{r_{\L\E}}+\frac32\frac{{\mu}_\S}{r^5_{\E\S}}\left[\left({\bm r}_{\E\S}\cdot{\bm r}_{\L\E}\right)^2-\frac13r_{\E\S}^2r_{\L\E}^2\right]\right\}dt
-\frac{1}{c^2}\Bigl({\bm v}_{\L\E}\cdot{\bm r}_{\E}-{\bm v}_{\L\E}\cdot{\bm r}_{\L\E}\Bigr)\;,                
\ea
which is in a full agreement with a physical intuition based on the principle of equivalence \citep{Ashby_2013} that tells us that local measurements cannot determine the speed of the orbital motion of a freely falling local coordinate system, and the gravitational field of external bodies can reveal itself solely in the form of the tidal terms. Indeed, the leading terms (the terms within the integral) of the TCL--TCG time transformation \eqref{11} are now clearly expressed solely in terms of the geocentric (relative) distance ${\bm r}_{\L\E}$ and velocity ${\bm v}_{\L\E}$ of the Moon with respect to the Earth and the terms depending on the barycentric orbital velocity, ${\bm v}_\E$, of the Earth cancel out.

It is worth noting that Eq. \eqref{11} can also be derived by introducing the local coordinates of the Earth-Moon system and matching them to the geocentric and selenocentric coordinates as described in Appendix \ref{appendix}. {This approach reveals the reason for the ``asymmetric'' appearance of the gravitational potentials of the Earth and Moon in the integral of Eq. \eqref{11}. The factor of 2 in the term $-2\mu_\L/r_{\L\E}$ arises from using the GCRS for calculating the TCL--TCG time transformation. When using the Earth-Moon local coordinate system described in Appendix \ref{appendix}, with the origin at the center of mass of the Earth-Moon system, the TCL--TCG transformation formula becomes fully symmetric with respect to the two bodies --- Earth and Moon --- as demonstrated in Eqs. \eqref{ap10} and \eqref{ap14}. Translating the origin of the Earth-Moon system to the geocenter, together with our preference for operating with relative distances and velocities referred to the GCRS, disrupts the symmetry of the time transformation formula, as shown in Eqs. \eqref{ap15}–\eqref{ap17}. This disruption introduces asymmetry in the appearance of the gravitational potentials of the Moon and Earth in Eq. \eqref{ap18}, which is equivalent to Eq. \eqref{a8}, a consequence of Eq. \eqref{11}.}

TCG represents the coordinate time of the local Geocentric Celestial Reference System (GCRS) and TCL represents the coordinate time of the local Lunar Celestial Reference System (LCRS).  Neither TCG nor TCL are the proper times measured by real clocks. Nonetheless, the GCRS and LCRS are mathematically constructed such that the proper time measured by the E-clock {at any geographic position on the Earth differs from TCG solely by a constant rate of approximately $60$ $\mu$s/day, while the proper time measured by the L-clock at any selenographic position on the Moon differs from TCL by a constant rate of approximately $2.7$ $\mu$s/day. The proof of this statement is given below and summarized in the form of Eqs. \eqref{a91a} - \eqref{11b}, \eqref{a8b} and \eqref{a8a}.}

Let us denote $\tau_\E$ as the proper time measured by a stationary\footnote{{A stationary position on Earth refers to a location that remains fixed relative to the Earth’s surface over time. A stationary clock is placed at a specific point defined by fixed latitude and longitude coordinates. This concept also applies to stationary clocks on the Moon. It is important to note that both Earth and Moon stationary clocks move relative to the non-rotating Geocentric Celestial Reference System (GCRS) and Lunar Celestial Reference System (LCRS), respectively, due to the rotation of the bodies, solid-body tides, and other effects.}} atomic clock on the Earth (the E-clock) and $\tau_\L$ as the proper time measured by a {stationary} atomic clock located at rest on the Moon (the L-clock). The world line of the E-clock is described in the non-rotating geocentric (GCRS) coordinate system by a time-dependent vector ${\bm w}={\bm w}(u)$. The transformation between the proper time $\tau_\E$ and the geocentric coordinate time $u=$TCG is given by equation \citep{kopeikin_2011book} 
\bal{a3}
\tau_\E=u-\frac{1}{c^2}\int\limits_{u_0}^u\left(\frac{{\dot w}^2}{2}+U_{\E}{+V_\E^{\rm tide}}\right)du\;,
\ea
where $\dot{\bm w}={\bm v}-{\bm v}_\E$ is the geocentric velocity of the E-clock with respect to the GCRS coordinates and $U_{\E}=U_\E({\bm w}(u))$ is the value of the gravitational potential of the Earth measured at the position of the E-clock, {and $V_\E^{\rm tide}=V_\E^{\rm tide}({\bm w}(u))$ represents the tidal gravitational potential of external bodies with respect to Earth. We have included the tidal potential in \eqref{a3} to ensure our theoretical considerations are valid for optical lattice clocks, achieving a systematic uncertainty in frequency of the order of $2\times 10^{-18}$ or better. Such clocks have been recently developed by researchers at JILA and the University of Colorado, Boulder \citep{Bothwell_2019,JILAclocks}. The exact expression for the tidal potential $V_\E^{\rm tidal}$ can be found in \citep{Agnew2007,Mueller_2018SSR}. Numerical estimates of the contribution of the tidal potential to Eq. \eqref{a3} are provided by \citet{Qin_2020AJ}. The constant average value of the tides is primarily defined by the, so-called, Doodson's constant, ${\cal D}_{\E\A}$, for the Earth \citep{Agnew2007}
\bal{dod1} 
\langle V_\E^{\rm tide}\rangle&\simeq&\sum_{\A\neq\E}{\cal D}_{\E\A}=\sum_{\A\neq\E}\frac{3\mu_\A}{4}\frac{R_\E^2}{r^3_{\E\A}}\;,
\ea 
referred to a particular body A contributing to the tidal potential. For Earth, the tidal potentials of the Sun (S) and the Moon (L) are the most significant. Their contribution to the secular drift of the E-clock caused by the tide's average value is 
\bal{dod1a}
\frac{{\cal D}_{\E\S}}{c^2} = 1.2\times 10^{-6}\;\; \mu{\rm s/day}\qquad,\qquad \frac{{\cal D}_{\E\L}}{c^2}\simeq 2.5\times 10^{-6}\;\; \mu{\rm s/day}\;.
\ea }\noindent
The integral in Eq. \eqref{a3} leads to a secular drift between the proper time of the E-clock and TCG. This secular drift is absorbed by rescaling TCG to the rate of the International Atomic Time (TAI) --- the latter based on the SI second on the rotating {\it geoid} --- but the conventional definitions of the two time scales retain the rate difference. 

Let the lunar L-clock have a fixed selenographic coordinates. The world line of the L-clock with respect to the selenocentric non-rotating (LCRS) coordinate system \footnote{The correspondence between the selenographic coordinates and the LCRS is discussed below in section \ref{F3} in more detail.} is described by vector ${\bm z}={\bm z}(s)$. The transformation between the proper time $\tau_\L$ and the lunar coordinate time $s=$TCL is given by the equation 
\bal{a1}
\tau_\L=s-\frac{1}{c^2}\int\limits_{s_0}^s\left(\frac{{\dot z}^2}{2}+U_{\L}{+V_\L^{\rm tide}}\right)ds\;,                
\ea
where $\dot{\bm z}={\bm v}-{\bm v}_\L$ is the velocity of the L-clock with respect to the LCRS, $U_\L=U_\L({\bm z}(s))$ is the value of the gravitational potential of the Moon at the position of the clock, {and $V_\L^{\rm tide}=V_\L^{\rm tide}({\bm z}(s))$ represents the tidal gravitational potential of external bodies with respect to Moon which must be included in \eqref{a1} to ensure that our formalism is valid for optical lattice clocks placed on the Moon. The estimate of the average secular drift of the clock caused by the tidal potential is evaluated by making use of the Doodson constants for the Moon \citep{Agnew2007}
\bal{dod2} 
\langle V_\L^{\rm tide}\rangle&\simeq&\sum_{\A\neq\L}{\cal D}_{\L\A}=\sum_{\A\neq\L}\frac{3\mu_\A}{4}\frac{R_\L^2}{r^3_{\L\A}}\;,
\ea 
referred to a particular body A contributing to the tidal potential. For Moon, the tidal potentials of the Sun (S) and the Earth (E) are the most significant. Their contribution to the secular drift of the L-clock caused by the tides are 
\bal{dod2a}
\frac{{\cal D}_{\L\S}}{c^2} = 8.7\times 10^{-8}\;\; \mu{\rm s/day}\qquad,\qquad \frac{{\cal D}_{\L\E}}{c^2}\simeq 1.5\times 10^{-5}\;\; \mu{\rm s/day}\;.
\ea }\noindent
The integrand in Eq. \eqref{a1} is practically constant for the clock placed on the surface of the Moon, leading to a secular drift of the proper time of the L-clock on the Moon with respect to TCL. {This secular drift is absorbed by rescaling TCL to the rate of lunar time on the Moon's surface (specifically, LT; see next page), which is based on the SI second there. The SI second, defined by the fixed numerical value of the cesium-133 atom's radiation frequency, is physically the same everywhere \citep{BIPM2019}. Therefore, we assume that SI seconds anywhere on the Earth or the Moon are identical to a local observer within the uncertainty allowed by the technical realization of time.} In the case where the clock is on a spacecraft orbiting the Moon, the integral in Eq. \eqref{a1} may include, in addition to the secular term, also time-dependent terms if the orbit has a sensible eccentricity \citep{Qin_2020AJ}. 

The physical model of the time transformation between the proper time measured by an atomic clock on the Moon with respect to one on the Earth consists of three parts: (1) the transformation \eqref{10} between TCG and TCL; (2) the transformation \eqref{a3} between the proper time $\tau_\E$ of the clock on Earth and TCG; and (3) the transformation \eqref{a1} between the proper time $\tau_\L$ of the clock on Moon and TCL.

Let us assume that the two clocks are located at rest respectively on the physical surfaces of the Earth and Moon. When calculating the proper time measured by each clock, it is useful to combine the quadratic velocity-dependent term and the Newtonian gravity potential within the integral in Eqs. \eqref{a3} and \eqref{a1} into a single function, $\Phi$, defined at the location of each clock as follows:
\bal{a81} 
\Phi_\E&:=&\frac{{\dot w}^2}{2}+U_{\E}=\frac12\left({\bm\omega}_\E\times{\bm w}\right)^2+U_{\E}\;\\
\label{a82}
\Phi_\L&:=&\frac{{\dot z}^2}{2}+U_{\L}=\frac12\left({\bm\omega}_\L\times{\bm z}\right)^2+U_{\L}\;,
\ea
where ${\bm\omega}_\E$ and ${\bm\omega}_\L$ are the angular velocity of rotation of the Earth and Moon respectively.  
Functions $\Phi_\E$ and $\Phi_\L$ represent correspondingly the gravity potentials in the rigidly rotating frames of the Earth and Moon, and they have constant values for clocks at rest on the surface of the Earth and Moon. {Strictly speaking, this statement is valid only if the angular velocity of the rotation remains constant. Deviations of $\Phi_\E$ and $\Phi_\L$ are caused by irregularities such as precession, nutation, and pole wobble in the rotational rate and the tidal deformations of the Earth and Moon crusts \citep{Qin_2020AJ}. It has been shown \citep{fateev_2015} that the relative shift in frequency and time due to these effects can reach for the Earth a relative uncertainty of $5\times 10^{-16}$, which is larger than the tidal effects but still insignificant for current practice.} 

In general, the constant numerical value of $\Phi_\E$ is slightly different for clocks having different geographic coordinates because the physical surface of the Earth on continents does not coincide with the {\it geoid}.  The geoid represents a reference equipotential surface of the Earth that coincides with the surface of the ocean undisturbed by tides and atmospheric phenomena \citep{torge2012}. The same statement applies to the lunar gravity potential
$\Phi_{\L}$ at the locations of clocks having different selenographic coordinates. In the case of the Moon, the reference equipotential surface of the gravitational potential is referred to as the {\it selenoid} \citep{Hirt_2012EPSL}, although an internationally accepted definition and height with respect to lunar surface features has not yet been established.

The variation of the gravity potential from one geographic location of a clock to another can be represented by a Taylor series of $\Phi_\E$ with respect to the reference value $\Phi_{0\E}=6.26368560\times 10^7$ ${\rm m}^2{\rm s}^{-2}$ of the gravity potential of the geoid \citep{Luzum11}.  The first two terms in the expansion of the gravity potential $\Phi_\E$ are \citep{torge2012,Mueller_2018SSR}
\bal{a91} 
\Phi_\E&=&\Phi_{0\E}+g_\E H_\E\;
\ea
where $H_\E$ is the orthometric height of the terrestrial E-clock above the geoid and $g_\E$ is the value of the acceleration of gravity at the position of the clock. 
The variation of the lunar gravity potential is, similarly,
\ba
\label{a92}
\Phi_{\L}&=&\Phi_{0\L}+\mathit{g}_\L H_\L\;.
\ea
where $\Phi_{0\L}=2.822336927\times 10^6$ ${\rm m}^2{\rm s}^{-2}$ is the value of the potential of the selenoid given in Ref. \citep{Ardalan_2014CeMDA}, $H_\L$ is the normal height of the lunar L-clock above the surface of the selenoid, and $\mathit{g}_\L$ is the value of the acceleration of gravity at the position of the clock on the Moon.

It is convenient to introduce Terrestrial Time (TT) and Lunar Time (LT).  TT is described as an idealized form of International Atomic Time, TAI (apart from a constant offset), with a rate that is nominally given by the SI second on the rotating geoid \citep[Resolution A4, Recommendation IV, pp. 45-47]{iau1992}. LT is useful for the same reason as TT in that it represents an idealized form of what might be called Lunar Atomic Time (a concept not yet adopted by international scientific bodies), which would be based on the SI second on the rotating selenoid {which we assume to be identical with SI second on the geoid}.  TT and LT are defined by similar equations 
\bal{a91a} 
{\rm TT}&=&\left(1-L_\E\right) {\rm TCG}\;,\\
\label{a92b}
{\rm LT}&=&\left(1-L_\L\right) {\rm TCL}\;,
\ea
where the scaling constants \footnote{The scale constant $L_\E$ is denoted as $L_\G$  in IAU Resolutions \citep{Luzum11}. We prefer to use the notation $L_\E$ to conform with the entire set of the labels adopted in the present paper for indexing various astronomical bodies of the solar system.}
\bal{11a}
L_\E&=&\frac{\Phi_{0\E}}{c^2}=60.2147\;\mu\mbox{\rm s/day}\;,\\
\label{11b}
L_\L&=&\frac{\Phi_{0\L}}{c^2}=2.7128\;\mu\mbox{\rm s/day}\;,
\ea
define the secular drift of the rate of clocks placed on the surface of the geoid and selenoid with respect to the coordinate times TCG and TCL, respectively. The secular rates shown in Eq. \eqref{11a} and Eq. \eqref{11b} correspond to the following adopted numerical values of the scale factors: $L_\E = 6.969\,290\,134 \times 10^{-10}$ \citep{Luzum11} and $L_\L=3.14027\times 10^{-11}$ \citep{Ardalan_2014CeMDA}.  {Note that, by definition, TT and LT are themselves coordinate times.}

Let us simplify the notations of the vectors entering Eq. \eqref{a81} and denote ${\bm r}:={\bm r}_{\L\E}$, ${\bm r}':={\bm r}_{\E\S}$, ${\bm v}:={\bm v}_{\L\E}$, ${\bm v}':={\bm v}_{\E\S}$ (see Section \ref{noconastr} for further details). The terms standing outside of the integral in Eq. \eqref{10} can be rearranged by considering the point ${\bm x}$ as the point of physical location of the lunar L-clock,
\bal{a7a}
{\bm v}_{\L\E}\cdot{\bm r}_{\E}-{\bm v}_{\L\E}\cdot{\bm r}_{\L\E}={\bm v}\cdot\left({\bm x}-{\bm x}_\L\right)={\bm v}\cdot{\bm z}\;,
\ea
where ${\bm z}:= {\bm r}_{\L}={\bm x}-{\bm x}_\L$ is the selenocentric position of the lunar L-clock. We also note that the term depending on the gravitational parameter $\mu_\S$ of the Sun in the right-hand side of Eq. \eqref{11} is, actually, the tidal gravitational potential of the Sun perturbing the orbital motion of the Moon \citep[p. 57]{koval1967}:
\bal{pp45}
W&:=&\frac32\frac{{\mu}_\S}{r^5_{\E\S}}\left[\left({\bm r}_{\E\S}\cdot{\bm r}_{\L\E}\right)^2-\frac13r_{\E\S}^2r_{\L\E}^2\right]=\frac32\frac{\mu_\S}{r'}\left(\frac{r}{r'}\right)^2\left[({\bm n}'\cdot{\bm n})^2-\frac{1}3\right]\;,
\ea
where $r'=|{\bm r}'|=|{\bm r}_{\E\S}|$ and we have introduced two unit vectors ${\bm n}':={\bm r}_{\E\S}/|{\bm r}_{\E\S}|={\bm r}'/r'$ and ${\bm n}:={\bm r}_{\L\E}/|{\bm r}_{\L\E}|={\bm r}/r$.

Then, the {physical model of the time transformation between terrestrial and lunar clocks} takes on the following form:
\bal{a8}
{\rm LT}&=&\Bigl(1+L_\E-L_\L\Bigr){\rm TT}-\frac{1}{c^2}\int\limits_{t_0}^t\left(\frac{v^2}{2}+\frac{{\mu}_\E-2{\mu}_\L}{r}\right)dt-\frac1{c^2}\int\limits_{t_0}^t W dt-\frac{1}{c^2}{\bm v}\cdot{\bm z}\;,\\
\label{a8b}
\tau_\E&=&\left(1-\frac{1}{c^2}\mathit{g}_\E H_\E {+\frac{1}{c^2}\langle V^{\rm tide}_\E\rangle}\right){\rm TT}\;,\\
\label{a8a}
\tau_\L&=&\left(1-\frac{1}{c^2}\mathit{g}_\L H_\L {+\frac{1}{c^2}\langle V^{\rm tide}_\L\rangle}\right){\rm LT} \;.
\ea
In Eq. \eqref{a8}, we need to perform a time integration along the world lines of both the Earth and the Moon. Additionally, this equation also considers the position-dependent contribution of the clock. This contribution is represented as the final term in Eq. \eqref{a8}. It encapsulates the special relativistic effect of time dilation experienced by a moving clock that has been relocated from the origin of the Lunar Celestial Reference System (LCRS) to a different position, denoted as ${\bm z}$, in space. We discuss the analytic results of integration of Eq. \eqref{a8} in next two sections. A numerical evaluation of the integral in Eq. \eqref{a8} along with discussion of other aspects of LT--TT time transformation are given in Section \ref{s7}  

Eqs. \eqref{a8b} and \eqref{a8a} contain only secular terms, which are proportional to the gravity potential and its gradient and reveal themselves in the form of a linear drift of the proper times of the atomic clocks from the corresponding coordinate times. The main difference between the proper time
$\tau_\E$ and TCG is in the secular rate given by the scale constant $L_\E$, shown in Eq. \eqref{a91a}, which is part of the IAU definition of the time scale TT.  An additional secular drift of the proper time $\tau_\E$ with respect to TT is due to the elevation $H_\E$ of the clock above the level of the geoid. This rate is determined by the value of the normal acceleration of gravity $g_\E$ on geoid, which depends on the latitude of the clock according to Somigliana's formula \citep[Eq. 4.41a]{torge2012}. This dependence gives a negligibly small contribution to the rate of clock. Thus, after taking the standard value of the acceleration of gravity on sea level, $g_\E=9.81$ m s$^{-2}$ and evaluating the rate of the secular drift between TT and the proper time $\tau_\E$ of a clock at the elevation $H_\E$, we obtain  
\bal{qq9}
\frac{1}{c^2}g_\E H_\E&=&9.4\times 10^{-6}\left(\frac{H_\E}{1\,{\rm m}}\right)\, \mu{\rm s}/{\rm day}\;.
\ea  

A similar consideration applies in evaluating the relationship \eqref{a8a} between the proper time of the L-clock on Moon and TCL. The main effect is a secular drift of $\tau_\L$ with respect to TCL given by the numerical value of the scale factor $L_\L$ in Eq. \eqref{11b}.  The elevation $H_\L$ of the lunar L-clock above the equipotential surface of the selenoid yields the following secular drift of the clock:
\bal{qq12}
\frac{1}{c^2}g_\L H_\L&=&1.6\times 10^{-6}\left(\frac{H_\L}{1\,{\rm m}}\right)\, \mu{\rm s}/{\rm day}\;,
\ea  
where we have used the mean value of the acceleration of gravity on lunar surface, $g_\L=1.63$ m\,s$^{-2}$. {Notice that the numerical estimates \eqref{qq9} and \eqref{qq12} of the secular drift of clocks are comprarable to the tidal contributions \eqref{dod1a} and \eqref{dod2a}. This is why we have included the tides to the transformation equations \eqref{a8b}, \eqref{a8a} making them valid for ultra-precise time measurements.}

\section{Analytic Model of the Earth-Moon-Sun System}\label{s5}  

To integrate the LT--TT transformation analytically as shown in Eq. \eqref{a8}, we need to express the integrand as a function of time. To do this, we use a simplified but realistic Keplerian model for the Earth-Moon-Sun system. This model overlooks perturbations caused by the major planets in the solar system {but renders analytic development of the theory possible}. We also assume that the barycenter of the Earth-Moon system aligns with the geocenter of the Earth when we calculate the tidal effects of the solar gravitational field on the Moon's motion relative to Earth. It is important to note that these simplifications are not applicable to more accurate numerical models, which we discuss in more detail in section \ref{s7}. 

{It is instructive to clarify why we have chosen to develop an analytic model of time transformations in this paper, which might seem unnecessary given the superiority of modern numerical computational methods in ephemeris astronomy. Although contemporary computers are indeed powerful and capable of solving equations numerically with great precision, there are several compelling reasons for pursuing the development of an analytic model.
\begin{itemize}
\item The analytic model provides deeper insights into the underlying physics and mathematics of the problem. It reveals relationships and dependencies that might not be immediately apparent from numerical solutions, such as the asymmetry in the appearance of the gravitational potentials in Eq. \eqref{11}.
\item The analytic solution makes it easier to understand and communicate the time transformation equations, highlighting key parameters and their effects, which are often obscured in numerical data. 
\item The analytic model serves as a benchmark to validate our numerical methods, ensuring that the numerical solution in Section \ref{s7} is accurate and reliable.
\item Our approximate analytic model can be readily generalized to a broader range of conditions and scenarios, such as clocks placed at the Lagrange points or clocks in satellites orbiting the Moon. This provides greater applicability compared to a specific numerical solution.
\item We believe that the analytic model is invaluable in education, helping students and researchers develop a strong foundational understanding of the principles governing the relativistic theory of time scales and reference frames in the solar system.
\end{itemize}
In summary, while numerical methods are incredibly powerful, analytic models complement them by providing clarity, efficiency, and deeper understanding. Both approaches are essential in advancing science and engineering.}

\subsection{The orbital elements of the Earth-Moon-Sun system}\label{sub1}
\subsubsection{The Earth's orbit}\label{subsub1}
Let's introduce the ecliptic reference frame defined by a triad of unit vectors:
\bal{ser6}
{\bm e}_x=(1,0,0)\qquad,\qquad{\bm e}_y=(0,1,0)\qquad,\qquad{\bm e}_z=(0,0,1)\;,
\ea
where ${\bm e}_x$ points toward the true equinox at the epoch $t_0$, within the plane of the ecliptic, ${\bm e}_y$ also lies in the ecliptic plane, and ${\bm e}_z={\bm e}_z\times{\bm e}_y$ points toward the ecliptic pole. The Keplerian model approximates Earth's orbital motion around the Sun as an ellipse within the ecliptic plane. Specifically, the heliocentric radius vector of the Earth is given by
\bal{x1}
{\bm r}'&=&r'\Bigl({\bm e}_x\cos \phi'+{\bm e}_y\sin \phi'\Bigr)\;,\\\label{x2}
r'&=&\frac{a'(1-e'^2)}{1+e'\cos\phi'}\;,
\ea
where $a'$ and $e'$ are the semi-major axis and eccentricity of the Earth's orbit, and $\phi'$ corresponds to the true anomaly of Earth's angular motion around the Sun. Our analytic model simplifies Earth's orbital dynamics by neglecting perturbations from the major planets. Consequently,
\begin{itemize}
\item the orbital elements $a'$ (semi-major axis) and $e'$ (eccentricity), which define Earth's orbit shape, remain constant;
\item the spatial orientation of Earth's orbit remains fixed: its inclination ($i'$) is zero, the pericenter ($\omega'$) is zero, and the ascending node
($\Omega'$) is zero. Notably, the angles $i'$, $\omega'$, and $\Omega'$ do not play a role in our calculations.
\end{itemize}
Moving forward, we adopt a linearized model with respect to eccentricity $e'$ for Earth's orbital motion. In this approximation, we calculate the true anomaly $\phi'$ in terms of the mean anomaly $M'$ of the Earth by using equation \citep[Eq. 35]{koval1967}
\bal{phiM}
\phi'=M'+2e'\sin M'\;.
\ea
Here, $M'$ varies linearly with time,
\bal{bi4} 
M'&=&M'_0+n'(t-t_0)\;,
\ea 
where the quantity $M'_0$ is the mean anomaly of the Earth at the epoch $t_0$,  and 
\bal{b5iu}
n'&=&\mu_\S^{1/2}{a'}^{-3/2}\;,
\ea
represents the constant angular rate of the mean orbital motion of the Earth around the Sun.

\subsubsection{The Moon's orbit}\label{subsub2}

The reference frame used for describing the orbital motion of the Moon around the Earth is defined by the triad of unit vectors placed at the center of mass of the Earth (geocenter). They are:
\bal{tr4q}
{\bm l}&=&\phantom{+}{\bm e}_x\cos\Omega+{\bm e}_y\sin\Omega \;,\\\label{tr5q}
{\bm m}&=&-{\bm e}_x\cos i\sin\Omega+{\bm e}_y\cos i\cos\Omega+{\bm e}_z\sin i \;,\\\label{tr6q}
{\bm k}&=&\phantom{+}{\bm e}_x\sin i\sin\Omega-{\bm e}_y\sin i\cos\Omega+{\bm e}_z\cos i \;,
\ea
where ${\bm l}$ points toward the ascending node of the lunar orbit, ${\bm m}$ lies in the plane of the lunar orbit, and ${\bm k}={\bm l}\times{\bm m}$ is orthogonal to the plane of the lunar orbit so that the orbital motion of the Moon is counterclockwise as seen from the tip of the vector ${\bm k}$. These vectors provide a convenient reference frame for analyzing the Moon's motion relative to the Earth. They assist in describing the orientation and geometry of the lunar orbit.

Let's consider the details of the Moon's orbital motion around the Earth using the Keplerian approximation. The Earth is situated at one of the foci of the Moon's elliptical orbit, which is also the origin of the geocentric coordinate reference system (GCRS) associated with the triad of vectors ${\bm l}$, ${\bm m}$ and ${\bm k}$. The key vectors defining the Moon's orbit are the radius vector ${\bm r}$ and velocity of the Moon ${\bm v}=\dot{\bm r}$. They are defined by the following equations \citep{brum}
\ba
\label{df5}
{\bm r}&=&r\Bigl({\bm P}\cos \phi+{\bm Q}\sin \phi\Bigr)\;,\\
\label{pm9}
{\bm v}&=&\sqrt{\frac{\mu_\E+\mu_\L}{a\left(1-e^2\right)}}\Bigl[-{\bm P}\sin \phi+{\bm Q}\left(\cos \phi+e\right)\Bigr]\;,
\ea
where $r:=|{\bm r}|$,
\bal{bi1}
r&=&\frac{a(1-e^2)}{1+e\cos \phi}\;,
\ea
the parameter $a$ represents the semi-major axis and $e$ is eccentricity of the lunar orbit, and 
\bal{po87n}
{\bm P}={\bm l}\cos\omega+{\bm m}\sin\omega\qquad,\qquad{\bm Q}=-{\bm l}\sin\omega+{\bm m}\cos\omega\;,
\ea
are the unit vectors in the plane of the lunar orbit. Vector ${\bm P}$ points  along the semi-major axis toward the pericenter, ${\bm Q}$ points along the semi-minor axes. These vectors depend on the angles $\omega$, the angular distance of the pericenter from the ascending node, and $\phi$, the true anomaly, which is a crucial parameter in describing the position of the Moon in its elliptical orbit. 

Let's express the sine and cosine of the true anomaly $\phi$ as functions of time using Fourier series with respect to the mean anomaly $M$ of the Moon \citep[p. 49]{koval1967}
\bal{bi3}
\sin\phi&=&\left(1-e^2\right)\sum_{k=1}^\infty \Bigl[J_{k-1}\left(ke\right)-J_{k+1}\left(ke\right)\Bigr]\sin kM\;,\\
\label{bi8}
\cos\phi&=&-e+2\frac{1-e^2}{e}\sum_{k=1}^\infty J_k\left(ke\right)\cos kM\;.
\ea
where the coefficients $J_k(ke)$ are the Bessel functions depending on the eccentricity of the lunar orbit. The Fourier series \eqref{bi3}, \eqref{bi8} can be further expanded with respect to the eccentricity by using a Taylor expansion of the Bessel functions
\bal{bv67m}
J_k(ke)=\sum_{p=0}^\infty\frac{(-1)^p}{p!(p+k)!}\left(\frac{ke}{2}\right)^{2p+k}\;.
\ea
In the Keplerian approximation, the mean anomaly of the Moon $M$ is an angular parameter that increases uniformly with time,  
\bal{123bi} 
M&=&M_0+n(t-t_0)\;,
\ea 
where $M_0$ is the mean anomaly of the Moon at the epoch $t_0$, and 
\bal{zxb5}
n&=&\left(\mu_\E+\mu_\L\right)^{1/2}a^{-3/2}\;,
\ea
represents the constant angular rate of the mean orbital motion of the Moon around the Earth.

\subsection{The perturbing potential}\label{sub3}

The effects of solar tidal perturbations emerge in the LT--TT time transformation \eqref{a8} in two ways: 
\begin{enumerate}
\item the explicit effects of the perturbing potential $W$, which accounts for the gravitational influence of the Sun on the Earth-Moon system and contributes directly to the transformation between lunar and terrestrial times,
\item the implicit effects from the perturbations of the Moon's radial distance $r$ and orbital velocity ${\bm v}$ due to solar tides. These perturbations appear under the integral sign in the transformation Eq. \eqref{a8}. The integral represents the cumulative effect of these small variations over time. 
\end{enumerate}

To thoroughly investigate the impact of both types of perturbations on the LT--TT time transformation, it is essential to provide a detailed analytical exposition of the perturbing potential $W$.
We begin this analysis with Eq. \eqref{pp45}, which defines $W$. This potential depends on the distances $r$ and $r'$ as well as the dot product ${\bm n}\cdot{\bm n}'$ of two unit vectors ${\bm n}$ and ${\bm n}'$. Making use of Eqs. \eqref{x1} and \eqref{df5}, we find that
\bal{x4}
{\bm n}'\cdot{\bm n}&=&\cos(\phi' - \Omega) \cos(\phi + \omega) + \cos i \sin(\phi' - \Omega)\sin(\phi + \omega)\;,
\ea
which is used for making $W$ an {explicit} function of time in Eq. \eqref{a8}. Unfortunately, this form of $W$ is not amenable to analytic integration due to its intricate functional dependence on time through the true anomalies $\phi$ and $\phi'$. In order to facilitate analytic calculations, we propose expanding the perturbing potential $W$ in a Taylor series with respect to several small parameters: the eccentricity $e\simeq 0.055$ of the Moon's orbit, the eccentricity $e'\simeq 0.017$ of the Earth's orbit, the inclination $i\simeq 0.090$ rad, and the ratio of orbital frequencies $n'/n\simeq 0.075$. By doing so, we can approximate the effects of these parameters on the LT--TT time transformation. The magnitudes of various terms arising from the expansion and integration of the direct and indirect perturbations in Eq. \eqref{a8} are shown in Table~1 {and their analytic expressions are present in Eqs. \eqref{numc4}--\eqref{numc4fin} and \eqref{x8}--\eqref{x8fin}.}
\vspace{0.2cm}
\begin{center}
    \begin{tabular}{|c|c|c|c|c|c|}
    \multicolumn{6}{c}{\bf Table 1. Magnitude of Periodic Terms in the LT--TT Time Transformation.} \\[0.2in] 
    \hline
        &&&&&\\
 $\phantom{\displaystyle\frac{n'a^2}{c^2}}$&$\displaystyle\frac{na^2}{c^2}$& $\displaystyle\frac{n'a^2}{c^2}$ & $\displaystyle\frac{n'^2a^2}{nc^2}$& $\displaystyle\frac{n'^3a^2}{n^2c^2}$&$\displaystyle\frac{n'^4a^2}{n^3c^2} $ \\
        &&&&&\\ \hline
        &&&&&\\
~~~~~~1~~~~~~&~~~4.38382~~~&~~~0.32792~~~&~~~0.02453~~~&~~~0.00183~~~&~~~0.00014~~~\\
        &&&&&\\\hline
        &&&&&\\
~~~~~~$e$~~~~~~&~~~0.24068~~~&~~~0.01800~~~&~~~0.00135~~~&~~~0.00010~~~&~$\sim 10^{-6}$~~~\\
        &&&&&\\\hline
        &&&&&\\
~~~~~~$e'$~~~~~~&~~~0.07321~~~&~~~0.00548~~~&~~~0.00041~~~&~~~0.00003~~~&~$\sim 10^{-6}$~~~\\
        &&&&&\\\hline  
        &&&&&\\
~~~~~~$i^2$~~~~~~&~~~0.03535~~~&~~~0.00264~~~&~~~0.00020~~~&~~~0.00002~~~&~$\sim 10^{-6}$~~~\\
        &&&&&\\\hline      
        &&&&&\\
~~~~~~~$e^2$~~~~~~~&~~~0.01321~~~&~~~0.00099~~~&~~~0.00007~~~&~$\sim 10^{-6}$~~~&~$\sim 10^{-7}$~~~\\
        &&&&&\\\hline
  &&&&&\\
~~~~~~$ee'$~~~~~~&~~~0.00402~~~&~~~0.00030~~~&~~~0.00002~~~&~$\sim 10^{-6}$~~~&~$\sim 10^{-7}$~~~\\
        &&&&&\\\hline              
        &&&&&\\
~~~~~~$e'^2$~~~~~~&~~~0.00122~~~&~~~0.00009~~~&~$\sim 10^{-6}$~~~&~$\sim 10^{-7}$~~~&~$\sim 10^{-8}$~~~\\
        &&&&&\\\hline             
    \end{tabular}
    \\[0.15in]
\parbox{5.4in}{\small {The table presents the numerical magnitudes of various terms in the analytic model of the LT-TT time transformation \eqref{a8} within the integral $c^{-2}\int Wdt$, following the expansion of the perturbing potential $W$ with respect to small parameters. More precise analytic expressions for these terms are provided in Eqs. \eqref{numc4}--\eqref{numc4fin} and \eqref{x8}--\eqref{x8fin}. The first column lists terms proportional to the eccentricities of the Moon’s ($e$)and Earth's ($e'$) orbits, as well as the angle of inclination ($i$) of the lunar orbit relative to the ecliptic. The second column, proportional to the mean motion $n$ of the Moon, indicates the magnitudes of the temporal periodic variations due to the Moon’s Keplerian motion. The remaining columns, proportional to powers of the Earth’s mean motion $n'$, show the magnitudes of the temporal periodic variations resulting from the direct and indirect tidal perturbations of the Moon’s Keplerian orbit by the Sun. All values are obtained by multiplying a term from the top row expression with the corresponding term from the first (left) column, and are given in microseconds ($\mu$s). }}
\end{center}

The table {demonstrates} that if we constrain the accuracy of our analytical calculation for the LT--TT time transformation to terms of the order of a few ns (1 ns $=$ 0.001 $\mu$s), we can simplify the expansion of the perturbing potential $W$ by retaining only the terms linear in $e'$ and quadratic in $e$ and $i$. By focusing on these specific terms, we strike a balance between accuracy and computational complexity. 
   
Let's provide a clearer presentation of the expansion of the perturbing potential $W$. We begin by expanding Eq. \eqref{x2} for $r'$ with respect to the small eccentricity $e'$, retaining only the linear term. Additionally, we expand \eqref{x4} in a series with respect to the inclination of the lunar orbit to the ecliptic, a small angle, keeping only the quadratic terms in $i$. Combining these expansions, we arrive at the following expression for the perturbing potential $W$:
\bal{x5}
W&=&n'^2r^2\left(1+3e'\cos\phi'\right)\left[\frac32\left(1-\frac{i^2}{2}\right)\cos^2\left(\phi + \omega+\Omega-\phi'\right)+\frac{3i^2}{8}\cos(2\phi+2\omega)+\frac{3i^2}{8}\cos(2\Omega-2\phi')-\frac12\right].
\ea 
Here, $n'$ represents the mean motion of the Earth around the Sun, as defined in Eq. \eqref{b5iu}.

We proceed by expanding the radial distance $r$ and the true anomaly $\phi$ of the lunar orbit in a series with respect to the eccentricity $e$ of the lunar orbit. We also expand Eq. \eqref{x5} in a Taylor series by expanding the true anomaly $\phi'$ with respect to the eccentricity of the orbital motion of the Earth-Moon system around the Sun. We utilize the expansions given in Eqs. \eqref{bi3} and \eqref{bi8} for the Moon's true anomaly $\phi$, and similar equations for the Earth's true anomaly $\phi'$ (replacing $e\rightarrow e'$) in the right-hand side of Eq. \eqref{x5}, and keep only the leading order terms with respect to the small parameters $e$, $e'$ and $i$. The resulting expression for the perturbing potential $W$ is (c.f. \citep[p. 99]{koval1967}):
\bal{perf7}
W&=&n'^2a^2\left[\frac14+\frac{3e^2}8-\frac{e}2\cos M-\frac{e^2}8\cos 2M+\frac{15e^2}{8}\cos\left(2D-2M\right)-\frac{9e}{4}\cos\left(2D-M\right)+\right.\\\nonumber
&&\left.\phantom{n'^2a^2}\left(\frac34-\frac{15e^2}{8}\right)\cos 2D+\frac{3e}{4}\cos\left(2D+M\right)+\frac{3e^2}{4}\cos\left(2D+2M\right)\right]\\\nonumber
&+&i^2n'^2a^2\left[-\frac38-\frac{3}{8}\cos 2D+\frac38\cos(2M+2\omega)+\frac38\cos(2D-2F)\right]\\\nonumber
&+&e'n'^2a^2\left[\frac{3}{4}\cos M'+\frac{21}{8}\cos(2D-M')-\frac38\cos(2D+M')\right]\\\nonumber
&+&\frac{3ee'}{4}n'^2a^2\left[-\cos(M+M')-\cos(M-M')-\frac12\cos(2D+M+M')\right.\\\nonumber
&&\phantom{\frac{3ee'}{4}n'^2a^2[}\left.+\frac72\cos(2D+M-M')+\frac32\cos(2D-M+M')-\frac{21}{2}\cos(2D-M-M')\right]
\;.
\ea
Here, $M'$ represents the mean anomaly of the Earth. The angles $D$ and $F$ are defined as
\bal{argD}
D&:=& M-M'+\omega+\Omega\;,\qquad\quad F:=M+\omega\;.
\ea 
The angle $D$ represents the Moon's mean elongation (angular distance of the mean longitude of the Moon from the mean longitude of the Sun) and the angle $F$ is the difference between the mean longitude of the Moon and the longitude of the node of the lunar orbit. The angles $D$, $F$, $M$ and $M'$ represent the canonical Delaunay variables \citep{Simon_1994}. 

The tidal potential \eqref{perf7} is expressed in terms of the trigonometric functions of the true anomalies $M'$ and $M$, which are linear functions of time (as given by Eqs. \eqref{bi4} and \eqref{123bi}) in the unperturbed Keplerian approximation. Hence, the above expression for $W$ can be easily integrated with respect to time after substitution of expression \eqref{perf7} into Eq. \eqref{a8} to determine the direct tidal contribution to the relationship between lunar and terrestrial times. Additionally, we need to account for the indirect tidal perturbation arising from the tidal deformation of the lunar orbit, which necessitates evaluating the tidal evolution of its orbital elements.

\subsection{Evolution of the orbital elements}\label{sb4}
The evolution of the orbital elements of the lunar orbit is determined by 
the Lagrange equations \citep[p. 41]{koval1967}. In particular, the evolution of semi-major axis $a$, eccentricity $e$, and the mean anomaly $M$ obey the following equations \citep{koval1967}:
\bal{nio9}
\frac{da}{dt}&=&\frac{2}{na}\frac{\pd W}{\pd M}\;,\\\label{nbrt6}
\frac{de}{dt}&=&\frac{1-e^2}{na^2e}\frac{\pd W}{\pd M}-\frac{\sqrt{1-e^2}}{na^2e}\frac{\pd R}{\pd \omega}\;,\\\label{jnyr4}
\frac{dM}{dt}&=&n-\frac{2}{na}\frac{\pd W}{\pd a}-\frac{1-e^2}{na^2e}\frac{\pd W}{\pd e}\;,
\ea
where $W$ is the perturbing potential \eqref{perf7}, and $n$ is the mean motion (mean angular rate of the orbital motion) of the Moon.  The mean
motion $n$ is related to the semi-major axis $a$ of the lunar orbit by Kepler's 3rd law, Eq. \eqref{zxb5}, where the semi-major axis $a$ is considered a function of time because of the tidal perturbation. Therefore, in order to calculate perturbation of the mean anomaly $M$, we need, first, to solve the Langrange equation \eqref{nio9} for $a$ and only after that proceed with the integration of Eq. \eqref{jnyr4}.

After taking $W$ from Eq. \eqref{perf7}, computing the partial derivatives, and reducing similar terms, Eqs. \eqref{nio9} and \eqref{nbrt6} take on the following forms:
\bal{cse5}
\frac{da}{dt}&=&-3\frac{an'^2}{n}\left[
\left(1-\frac{i^2}{2}\right)
\sin {2D}-\frac{3e}{2}\sin(2D-M)+\frac{3e}{2}\sin(2D+M)\right.\\\nonumber
&&\left.  +\frac{7e'}{2}\sin(2D-M')-\frac{e'}{2}\sin(2D+M')-\frac{e}{3}\sin M
\right]\;,
\\\label{hu8o}
\frac{de}{dt}
&=&\frac{{n}'^2}{n}\left[\frac{1}2\sin M+\frac{e}4\sin 2M-\frac{9}{4}\sin\left(2D-M\right)-\frac{3}{4}\sin\left(2D+M\right)\right.\\\nonumber
&&\left.~~~+\frac{15e}{4}\sin\left(2D-2M\right)+\frac{3e}{4}\sin {2D}-\frac{3e}{2}\sin\left(2D+2M\right)\right]\\\nonumber
&+&\frac{3e'}{4}\frac{n'^2}{n}\left[\sin(M+M')+\sin(M-M')+\frac12\sin(2D+M+M')\right.\\\nonumber
&&\phantom{\frac{3ee'}{4}}\left.-\frac72\sin(2D+M-M')+\frac32\sin(2D-M+M')-\frac{21}{2}\sin(2D-M-M')\right]\;,
\ea
where we have retained only the terms which amplitude exceeds 0.1 ns in LT--TT time transformation. The solutions of these equations are
\bal{hg6o}
a=\a+\delta a\qquad\qquad,\qquad\qquad e=\e+\delta e\;,
\ea
where $\a$, $\e$ are constants, and $\delta a$, $\delta e$ are perturbations. The perturbations are obtained by direct integration of Eqs. \eqref{cse5} and \eqref{hu8o} by employing the unperturbed mean anomalies \eqref{bi4}, \eqref{123bi} with constant values of the mean motions \eqref{b5iu}, \eqref{zxb5}. 
We obtain
\ba\label{hgu2}
\delta a&=&\frac{3a}{2}\frac{n'^2}{n^2}\left[\left(1+\frac{n'}{n}-\frac{i^2}{2}\right)\cos 2D-3e\cos(2D-M)+e\cos(2D+M)\right.\\\nonumber
&&\left. \phantom{~~~~~~} +\frac{7e'}{2}\cos(2D-M')-\frac{e'}{2}\cos(2D+M')-\frac{2e}{3}\cos M
\right]\;,\\
\label{nbv6y}
\delta e&=&\frac{15e}{8}\frac{n'}{n}\cos(2D-2M)-e\frac{n'^2}{n^2}\left[\frac{1}8\cos 2M+\frac{3}{8}\cos 2D
-\frac{3}{8}\cos\left(2D+2M\right)\right]\\\nonumber
&-&\frac{e'}{4}\frac{n'^2}{n^2}\left[3\cos(M+M')+3\cos(M-M')+\frac12\cos(2D+M+M')\right.\\\nonumber
&&\phantom{\frac{3ee'}{4}}\left.-\frac72\cos(2D+M-M')+\frac92\cos(2D-M+M')-\frac{63}{2}\cos(2D-M-M')\right]
\\\nonumber
&+&\frac{n'^2}{n^2}\left[-\frac{1}2\cos M+\frac{9}{4}\left(1+\frac{2n'}{n}\right)\cos\left(2D-M\right)
+\frac{1}{4}\left(1+\frac{2n'}{3n}\right)\cos\left(2D+M\right)\right]
\;.
\ea

Substituting the perturbed value of $a=\a+\delta a$ into the expression \eqref{zxb5} for the mean motion $n$, and replacing this expression for $n$ in
Eq.~\eqref{jnyr4} for the perturbed mean anomaly, yields
\ba\label{fgt5}
\frac{dM}{dt}&=&{n}\left(1-\frac74\frac{{n}'^2}{{n}^2}\right)+\frac{{n}'^2}{{ne}}\left[\frac{1}{2}\cos M+\frac{9}{4}\cos(2D-M)-\frac{3}{4}\cos(2D+M)\right]\\\nonumber
&+&\frac{{n}'^2}{{4n}}\left[\cos 2M+3\cos{2D}-15\cos(2D-2M)-6\cos(2D+2M)-21e'\cos M'\right]\\\nonumber
&+&\frac{3e'}{4e}\frac{{n}'^2}{{n}}\left[\cos(M+M')+\cos(M-M')
+\frac12\cos(2D+M+M')\right.\\\nonumber
&&~~~~~~~\left.-\frac72\cos(2D+M-M')-\frac32\cos(2D-M+M')+\frac{21}{2}\cos(2D-M-M')\right]\;,
\ea
where $n'$ and $n$ on the right hand side are considered to have constant values defined by Eqs. \eqref{b5iu}, \eqref{zxb5}.

Integrating Eq. \eqref{fgt5} once with respect to time results in
\bal{tg67}
M&=&\mathsf{M}+\delta M\;,
\ea
where $\mathsf{M}$ is a secular perturbation 
\ba\label{as6g}
\mathsf{M}&=&M_0+n\left(1-\frac74\frac{n'^2}{n^2}\right)(t-t_0)\;,
\ea
and 
\ba\label{nub5}
\delta M&=&
\frac1{e}\frac{n'^2}{n^2}\left[\frac12\sin M+\frac94\left(1+\frac{2n'}{n}\right)\sin(2D-M)-\frac14\left(1+\frac{2n'}{3n}\right)\sin(2D+M)\right]\\\nonumber
&&+\frac{n'}{n}\left[\frac{15}{8}\sin(2D-2M)-\frac{21}{4}e'\sin M'\right]
+\frac{n'^2}{n^2}\Bigl[\frac18\sin 2M+\frac38\sin 2D-\frac38\sin(2D+2M)\Bigr]\\\nonumber
&+&\frac{e'}{e}\frac{{n}'^2}{{n}}\left[\frac34\sin(M+M')+\frac34\sin(M-M')
+\frac18\sin(2D+M+M')\right.\\\nonumber
&&~~~~~~~\left.-\frac78\sin(2D+M-M')-\frac98\sin(2D-M+M')+\frac{63}{8}\sin(2D-M-M')\right]\;,
\ea
are periodic perturbations of the mean anomaly. 
An important detail to note is that on the right-hand side of Eq. \eqref{nub5}, the eccentricity is present in the denominator. However, this does not result
in a degeneracy in the LT--TT time transformation because all terms depending on $\delta M$ appear as the product $e\,\delta M$. This ensures that the eccentricity in the denominator does not introduce any singularities or undefined values in the LT--TT time transformation. This is crucial for the maintenance of stability and accuracy in the transformation. 

We have conducted a comparison between our analytic computation of the primary periodic components of lunar osculating elements $a$, $e$, $M$ and the semi-analytic theory of lunar motion as proposed by \citet{Simon_1994}. The agreement between these two results is within a 5--10\% margin for the initial three terms in the expansions \eqref{hgu2}, \eqref{nbv6y} and \eqref{nub5}. However, the level of agreement deteriorates to approximately 50--60\% for the subsequent terms. The explanation lies in the relatively large value of the small parameter $n'/n\simeq 0.075$ used in the perturbation theory of the osculating elements of the lunar orbit. This parameter is significant enough that including additional terms in the perturbation series results in a substantial contribution to the numerical coefficients of the expansion. This underscores the importance of considering higher-order terms in the series for a more accurate analytic representation. 

The remaining three osculating elements of the lunar orbit --- the inclination of the orbit to the ecliptic $i$, the argument of the pericenter $\omega$, and the longitude of the ascending node $\Omega$  --- are also perturbed by the tidal gravitational field of the Sun. Equations for perturbations of these elements are \citep{koval1967}
\bal{os7}
\frac{di}{dt}&=&\frac{\cos i}{na^2\sqrt{1-e^2}}\frac{1}{\sin i}\frac{\pd W}{\pd \omega}-\frac{1}{na^2\sqrt{1-e^2}}\frac{1}{\sin i}\frac{\pd W}{\pd \Omega}\;,\\\label{os8}
\frac{d\omega}{dt}&=&\frac{\sqrt{1-e^2}}{na^2e}\frac{\pd R}{\pd e}-\frac{\cos i}{na^2\sqrt{1-e^2}}\frac{1}{\sin i}\frac{\pd W}{\pd i}\;,\\\label{os9}
\frac{d\Omega}{dt}&=&\frac{1}{na^2\sqrt{1-e^2}}\frac{1}{\sin i}\frac{\pd W}{\pd i}\;,
\ea
where $W$ is the perturbing potential \eqref{perf7}.

The orbital inclination $i$ appears explicitly only in the perturbing potential itself. Therefore, its variation affects the orbital elements only in the second-order approximation and can be ignored. Hence, we can consider the orbital inclination $i$ as constant. The periodic variations of the orbital elements
$\Omega$ and $\omega$ are also insignificant for the purposes of the present paper. However, their secular variations are important for the correct
prediction of the periods of the periodic terms in the LT--TT transformation. Direct computation of the secular variations of $\Omega$ and $\omega$ yields
\bal{omOm}
\frac{d\Omega}{dt}=-\frac34\frac{n'^2}{n}\left(1+\frac{e^2}2\right)=-20.16^\circ\;{\rm yr}^{-1}\qquad\quad,\quad\qquad \frac{d\omega}{dt}=\frac32\frac{n'^2}{n}=39.09^\circ\;{\rm yr}^{-1}\;,
\ea 
which gives periods 17.86 and 9.21 years, respectively. More precise values in the complete theory of the Moon's motion are 18.60 and 8.85 years \citep{Simon_1994}.

\section{Analytic Model of the LT--TT Time Transformation}\label{s5a}   

In this section we consider the analytic model for the LT--TT time transformation given by Eq. \eqref{a8}. We write this equation in the form of a linear superposition of four terms
\bal{qq1}
{\rm LT}-{\rm TT}&=&\mathcal{F}_0(t)+\mathcal{F}_1(t)+\mathcal{F}_2(t)+\mathcal{F}_3(t)\;.
\ea
Here, the first term on the right-hand side 
\bal{qq0}
\mathcal{F}_0(t)&=&\left(L_\E-L_\L\right){\rm TT}\;,
\ea
describes a secular drift of LT with respect to TT;  the second term, given by the integral
\bal{qq2}
\mathcal{F}_1(t)&=&-\frac{1}{c^2}\int\limits_{t_0}^t\left(\frac{v^2}{2}+\frac{{\mu}_\E-2{\mu}_\L}{r}\right)dt\;,
\ea
describes the quadratic Doppler and gravitational time dilation effects caused by the orbital motion of Moon, the gravitational field of the Earth-Moon system, and the indirect tidal perturbation of the lunar orbit by the Sun; the integral
\bal{qq3}
\mathcal{F}_2(t)&=&-\frac{1}{c^2}\int\limits_{t_0}^tW dt\;
\ea
describes the direct time dilation effect of the tidal gravitational field of the Sun; and the instantaneous time-dependent term 
\bal{qq4}
\mathcal{F}_3(t)&=&-\frac{1}{c^2}\,{\bm v}\cdot{\bm z}
\ea
describes a special-relativistic time dilation effect caused by the orbital and rotational motion of the lunar L-clock having the (time-dependent) LCRS 
position ${\bm z}={\bm z}(t)$. 

Analytic expressions for each term can be derived by applying the Keplerian orbital approximation, supplemented with the secular and periodic variations of the lunar orbital elements. These variations are induced by solar tidal perturbations, as outlined in section \ref{s5}. It is important to note that the time argument used in Eqs. \eqref{qq1}--\eqref{qq4} is TCB $=t$. {However, the difference between TCB and TT is of the post-Newtonian order of magnitude (rate
difference $\approx 1.55\times10^{-8}$), rendering its contribution negligible (of the post-post-Newtonian order) in the calculation of the integrals. Consequently, we can treat functions
$\mathcal{F}_1$, $\mathcal{F}_2$ and $\mathcal{F}_3$ as being expressed in the TT time scale.}

\subsection{Computing the integral \texorpdfstring{$\mathcal{F}_1$}{F1}}\label{2mn7}
The integral $\mathcal{F}_1(t)$ can be evaluated analytically by considering the lunar orbit as an osculating Keplerian ellipse with its focus at the geocenter and its semi-major axis $a$ and eccentricity $e$ given in Eq. \eqref{hg6o}. Making use of the osculating Keplerian orbit representation \eqref{bi1} -- \eqref{bv67m}, and expanding the Bessel functions in power series with respect to eccentricity $e$, yields
\bal{bi9}
\frac{v^2}{2}+\frac{{\mu}_\E-2{\mu}_\L}{r}&=&\frac{3{\mu}_\E}{2a}\left(1-\frac{\mu_\L}{\mu_\E}\right)+\frac{{\mu}_\E}{a}\left(2-\frac{\mu_\L}{\mu_\E}\right)\left[\left(1-\frac{e^2}{8}\right)e\cos M+e^2\cos 2M+\frac{9}{8}e^3\cos 3M\right]\;.
\ea
Coefficients of the series \eqref{bi9} depend on the osculating elements of the lunar orbit --- the semi-major axis $a$ and eccentricity $e$ that are subject to the tidal perturbations from the Sun. The true anomaly $M$ is also perturbed by the solar tide. These perturbations produce the indirect tidal deformation of function \eqref{bi9} that has the same order of magnitude as the direct tidal effects on the rate of lunar time caused by the potential $W$. Hence, the effects of the indirect perturbation must be taken into account and calculated explicitly. This can be done with the help of Eqs. \eqref{hgu2}, \eqref{nbv6y}, and \eqref{nub5} for the perturbations of the osculating elements of the lunar orbit. 

Using the perturbed values of the orbital elements, expanding the right-hand side of Eq. \eqref{bi9} in a Taylor series, and keeping only linear terms with respect to the perturbations, gives us
\bal{onyt8}
\mathcal{F}_1(t)&=&-n^2\int\limits_{t_0}^t\Bigl[f(t)+\delta f(t)\Bigr]dt\;,
\ea
where 
\bal{tev5}
f(t)&=&\alpha_1+\alpha_2\left[\left(1-\frac{e^2}{8}\right)e\cos 
M+e^2\cos 2M+\frac{9}{8}e^3\cos 3M\right]
\ea
is the time series representation of the Doppler and gravitational shift effects, while
\bal{nt6v4}
\delta f(t)&=&-\alpha_1\frac{\delta a}{a}+\alpha_2\left[\left(\delta e-e\frac{\delta a}{a}\right)\cos M +2e\delta e\cos 2M-e\delta M\sin M-2e^2\delta M\sin 2M\right]
\ea
is the time series representation of the effect of the deformation of the lunar orbit caused by the tidal gravitational field of the Sun.
The constant numerical coefficients in Eqs. \eqref{tev5}, \eqref{nt6v4} are
\bal{yc5d}
\alpha_1=\frac{3}{2}\frac{\mu_\E-\mu_\L}{\mu_\E+\mu_\L}\frac{a^2}{c^2}\quad\qquad,\quad\qquad\alpha_2=\frac{2\mu_\E-\mu_\L}{\mu_\E+\mu_\L}\frac{a^2}{c^2}\;,
\ea
and the mean anomaly $M$ is given by Eq. \eqref{tg67}.

We further use Eqs. \eqref{hgu2}, \eqref{nbv6y}, and \eqref{as6g} to calculate the function $\delta f(t)$ explicitly. It results in\newline
\bal{ygb4r}
\delta f(t)&=&-\frac{\alpha_2}2\frac{n'^2}{n^2} + e\frac{n'^2}{n^2}\left({\alpha_1} - \frac98 {\alpha_2} \right)\cos M + 
 e\frac{n'}{n}\left[\frac{15}{8}{\alpha_2}+\left(\frac92 {\alpha_1}  - 
 \frac{1}{16} {\alpha_2} \right)\frac{n'}{n}\right] \cos(2D-M)  \\\nonumber
&-&\frac{n'^2}{n^2}\left(\frac32 {\alpha_1} - \frac52 {\alpha_2}\right)  \cos 2D- 
 e\frac{n'^2}{n^2}\left(\frac32{\alpha_1} - 
 \frac{57}{16} {\alpha_2} \right) \cos(2D+M)-\frac{33}8\frac{n'^2}{n^2}\alpha_2e^2\cos(2D-2M)\\\nonumber
&-& 
 e'\frac{n'^2}{n^2}\left[\left(\frac{21}4{\alpha_1} - 
 \frac{35}{4} {\alpha_2} \right) \cos(2D-M)-\left(\frac{3}4{\alpha_1} - 
 \frac{5}{4} {\alpha_2} \right) \cos(2D+M)+\frac32\alpha_2\cos M'\right]\\\nonumber
&+&\alpha_2e'e\frac{n'}{n}\left[\frac{21}8\cos(M-M')-\frac{21}8\cos(M+M')+\frac{35}8\cos(2D-M-M')-\frac{15}8\cos(2D-M+M')\right] \;.\\\nonumber
\ea
Integrating functions $f(t)$ and $\delta f(t)$ in Eq. \eqref{onyt8} by making use of Eqs. \eqref{bi4} and \eqref{123bi} for the unperturbed mean anomalies $M'$ and $M$, respectively, yields
\bal{st56h}
\mathcal{F}_1(t)&=&A_0(t-t_0)+A_1\sin M+A_2\sin 2M+A_3\sin 3M+A_4\sin(2D-M)+A_5\sin 2D
\\\nonumber
&+&A_6\sin(2D+M)+A_7\sin(M-M')+A_8\sin M'+A_9\sin(M+M')+A_{10}\sin(2D-2M)
\\\nonumber
&+&A_{11}\sin(2D-M')+A_{12}\sin(2D+M')+A_{13}\sin(2D-M-M')+A_{14}\sin(2D-M+M')\;,
\ea
where the analytic expressions for the numerical coefficients are given below:
\begin{subequations}
\ba
\label{numc4}
A_0&=&-\alpha_1n^2+\frac12\alpha_2n'^2\;,\\
A_1&=&-en\left[\alpha_2\left(1-\frac{e^2}{8}\right)+\left(\alpha_1-\frac98\alpha_2\right)\frac{n'^2}{n^2}\right]\;,\\
A_2&=&-\frac12\alpha_2e^2n\;,\\
A_3&=&-\frac38\alpha_2e^3n\;,\\
A_4&=&+en'\left[-\frac{15}{8}{\alpha_2}-\left(\frac92 {\alpha_1}  + 
 \frac{59}{16} {\alpha_2} \right)\frac{n'}{n}\right]\;,\\
A_5&=&+\frac{n'^2}{n}\left(\frac34 {\alpha_1} - \frac54 {\alpha_2}\right)\;,\\
A_6&=&+e\frac{n'^2}{n}\left(\frac12{\alpha_1} - 
 \frac{19}{16} {\alpha_2}\right)\\
A_7&=&-\frac{21}{8}\alpha_2ee'n'\;,\\
A_8&=&+\frac32\alpha_2 e'n'\;,\\
A_9&=&+\frac{21}{8}\alpha_2ee'n'\;,\\
A_{10}&=&+\frac{33}{16} \alpha_2 e^2 n'\;,\\
A_{11}&=&+e'\frac{n'^2}{n}\left(\frac{21}{8}{\alpha_1} - 
 \frac{35}{8} {\alpha_2}\right)\;,\\
A_{12}&=&-e'\frac{n'^2}{n}\left(\frac{3}{8}{\alpha_1} - 
 \frac{5}{8} {\alpha_2}\right)\;,\\
A_{13}&=&-\frac{35}{8}\alpha_2ee'n'\;,\\
\label{numc4fin}
A_{14}&=&+\frac{15}{8}\alpha_2ee'n'\;. 
\ea
\end{subequations}
Notice that each of these coefficients is actually represented in the form of an infinite power series with respect to the small parameters $e$, $e'$, and $n'/n$.  The above equations provide only the leading terms of the series in the expansion of the numerical coefficients, which is sufficient for our analytic study. 

\subsection{Computing the integral \texorpdfstring{$\mathcal{F}_2$}{F2}}\label{subF2}
The integral in Eq. \eqref{qq3} defines the tidal time dilation $\mathcal{F}_2(t)$ that can be computed by direct integration of the perturbing potential $W$ given by Eq. \eqref{perf7}. Substituting Eq. \eqref{perf7} in Eq. \eqref{qq3} and integrating with respect to time yields the following result:
\bal{x7}
\mathcal{F}_2(t)&=&B_0(t-t_0)+B_1\sin M+B_2\sin 2M+B_3\sin 3M+B_4\sin M'\\\nonumber
&+&B_5\sin(2M'-2\Omega)
+B_6\sin(2D-M) +B_7\sin 2D+B_8\sin(2D+M)\\\nonumber
&+&B_9\sin(2D-2M)
+B_{10}\sin(2D-M')+B_{11}\sin(2D+M')\;,
\ea
where the analytic expressions for constant coefficients are 
\begin{subequations}        
\bal{x8}
B_0&=&-\frac14\frac{n'^2a^2}{c^2}\left(1+\frac{3e^2}2-\frac{3i^2}2\right)\;,\\
B_1&=&+\frac{e}{2}\frac{n'^2a^2}{n c^2}\;,\\
B_2&=&+\frac{e^2}{16}\frac{n'^2a^2}{n c^2}\;,\\
B_3&=&-\frac{e^3}{16}\frac{n'^2a^2}{n c^2}\;,\\
B_4&=&-\frac{3e'}{4}\frac{n'a^2}{c^2}\;,\\
B_5&=&-\frac{3i^2}{16}\frac{n'a^2}{c^2}\;,\\
B_6&=&+\frac{9e}{4}\frac{n'^2a^2}{n c^2}\;,\\
B_7&=&-\frac{3}{8}\frac{n'^2a^2}{n c^2}\;,\\
B_8&=&-\frac{e}{4}\frac{n'^2a^2}{n c^2}\;,\\
B_9&=&\frac{15e^2}{16}\frac{n'a^2}{c^2}\;\\
B_{10}&=&-\frac{21e'}{16}\frac{n'^2a^2}{nc^2}\;\\
\label{x8fin}
B_{11}&=&\frac{3e'}{16}\frac{n'^2a^2}{nc^2}\;.
\ea
\end{subequations}
It is evident that the rate $B_0$ of the secular drift of TCL with respect to TCG, caused by the tidal gravitational potential of the Sun, is substantially smaller (by a factor $n'^2/n^2\simeq 0.005$) than the rate $A_0$ of the secular drift of function $\mathcal{F}_1$. Nonetheless, the amplitudes of the periodic terms in Eqs. \eqref{st56h} and \eqref{x7} for
$\mathcal{F}_1$ and $\mathcal{F}_2$ are comparable, even though many terms exhibit distinct frequencies. 

Making use of the numerical values of the astronomical constants for the Earth-Moon-Sun system, the analytic model gives the following numerical estimate for the sum of the two functions:\newline
\bal{pet52}
\mathcal{F}_1(t)+\mathcal{F}_2(t)&=&C_0(t-t_0)+C_1 \sin M +C_2 \sin 2M +C_3 \sin 3M
 +C_4 \sin(2D - M) +C_5 \sin 2D\\\nonumber
 &+& C_6 \sin(2D + M)+ C_7\sin M' +C_8 \sin(2D-2F) + 
 C_9 \sin(2D-2M)\\\nonumber
 &+&  C_{10}\sin(2D - M') +C_{11} \sin(2D + M')  +C_{12}\sin(M - M') +C_{13} \sin(M + M')\\\nonumber
 &+&C_{14}\sin(2D-M+M')+C_{15}\sin(2D-M-M')\;.
 \ea
where the angle $F:=M+\omega$.  The coefficient $C_0$ defines the secular rate of TCL with respect to TCG. Its numerical value is 
\bal{bh4d}
C_0&=&-1.4714\;\, \mu{\rm s}/{\rm day}\;.
\ea
The amplitudes of the periodic terms in Eq. \eqref{pet52} are 
\ba \nonumber
C_1&=&-0.4707 \;\,\mu{\rm s}\;,\qquad
C_2=- 0.0130 \;\,\mu{\rm s}\;,\qquad
C_3=- 0.0005 \;\,\mu{\rm s}\;,\qquad
C_4=- 0.0814 \;\,\mu{\rm s}\;,\\\nonumber
C_5&=&- 0.0468 \;\,\mu{\rm s}\;,\qquad
C_6=- 0.0025 \;\,\mu{\rm s}\;,\qquad
C_7=+ 0.0120\;\,\mu{\rm s}\;,\qquad
C_8=+ 0.0005 \;\,\mu{\rm s}\;,\\\nonumber
C_9&=&-0.0031 \;\,\mu{\rm s}\;,\quad~\;\;\,
C_{10}=-0.0024 \;\,\mu{\rm s}\;,\quad\;\;
C_{11}=+0.0003 \;\,\mu{\rm s}\;,\quad\;\;
C_{12}=- 0.0016 \;\,\mu{\rm s}\;,\\
C_{13}&=&+0.0016 \;\,\mu{\rm s}\;,\qquad
C_{14}=+0.0011\;\,\mu{\rm s}\;,  \qquad
C_{15}=-0.0025\;\,\mu{\rm s}\;.
\label{num38}
\ea
                                                                                                                             
\subsection{Computing the term \texorpdfstring{$\mathcal{F}_3$}{F3}} \label{F3}     
The instantaneous periodic function $\mathcal{F}_3(t)$ defined in Eq. \eqref{qq4} can be calculated by using the orbital velocity of the Moon \eqref{pm9} and the LCRS coordinates ${\bm z}$ of the L-clock on the lunar surface. The clock coordinates are introduced in three steps. First, we introduce the lunar coordinates associated with the mean rotational motion of the Moon, with the origin at the center of mass of the Moon. They are defined by the triad of unit vectors $\left\{{\bm a}, {\bm b}, {\bm c}\right\}$ which obey the empirical Cassini laws of the mean rotational motion of the Moon:
\begin{itemize}
\item[(1)] the Moon's rotation axis ${\bm c}$, its orbit normal ${\bm k}$, and the normal to the ecliptic ${\bm e}_3$ are coplanar,
\bal{g1}
\sin i\left({\bm c}\times{\bm e}_3\right)&=&\sin I\left({\bm e}_3\times{\bm k}\right)\;,
\ea
where $i$ and $I$ are the small angles of the inclination of the lunar orbit and the lunar equator to ecliptic;
\item[(2)] the ascending node of Moon's mean equator is always opposite the ascending node of Moon's orbit
\bal{g2}
{\bm e}_3\times{\bm c}&=&-(\sin I){\bm l}\;,
\ea
which means that the lunar equator precesses around the ecliptic at the same rate as Moon's orbit.
\end{itemize}
Here, the unit vectors ${\bm l}$ and ${\bm k}$ of the Moon's orbit are defined by Eqs. \eqref{tr4q}, \eqref{tr6q}.    

For the purpose of this paper, it suffices to disregard the small terms that are quadratic with respect to the inclination of the lunar equator to the ecliptic, $I={1^\circ}.542=0.027\; {\rm rad}\ll 1$. The triad of unit vectors $\left\{{\bm a}, {\bm b}, {\bm c}\right\}$ is rotating relative to the BCRS axes with a mean angular velocity that is equal to the mean orbital motion of the Moon. Therefore, accounting for Eqs. \eqref{g1}, \eqref{g2}, the triad $\left\{{\bm a}, {\bm b}, {\bm c}\right\}$  is expressed with respect to the ecliptic frame as follows \citep[Section 4.7]{Murray_book}:
\bal{n7bk}
{\bm a}&=&-{\bm e}_x\cos(M + \omega + \Omega) -{\bm e}_y\sin(M +\omega + \Omega)+{\bm e}_z\sin I\sin(M + \omega)\;,\\\label{htv6j}
{\bm b}&=&~~{\bm e}_x\sin(M + \omega + \Omega) -{\bm e}_y\cos(M +\omega + \Omega)+{\bm e}_z\sin I\cos(M + \omega)\;,\\\label{h6j22}
{\bm c}&=&-{\bm e}_x\sin I\sin\Omega +{\bm e}_y\sin I\cos\Omega+{\bm e}_z\;,
\ea
where $M$ is the mean orbital motion. Notice that vector ${\bm a}$ is almost anti-aligned with the radius-vector ${\bm r}$ of the Moon at any instant of time (c.f. Eqs. \eqref{df5} and \eqref{n7bk}).

As a second step, we introduce a triad of unit vectors $\left\{{\bm A}, {\bm B}, {\bm C}\right\}$ which is rigidly attached to the principal axes of Moon's figure, with vectors ${\bm A}$ and ${\bm C}$ directed along the axis of the minimal and maximal moments of inertia of the Moon, respectively, and ${\bm B}={\bm C}\times{\bm A}$. Finally, we consider the L-clock on the surface of the Moon, with a position defined by the vector ${\bm z}$ that is fixed with respect to selenographic coordinates.  The clock's selenographic coordinates are
\bal{sel8}
{\bm z}&=&\rho\left({\bm A}\cos b\cos l+{\bm B}\cos b\sin l+{\bm C}\sin b\right)\;,
\ea
where $\rho$ is the radial distance of the clock from the center of the Moon, and the angles $-90^\circ\le b\le +90^\circ$, $0\le l\le 360^\circ$ are the selenographic latitude and longitude, respectively, with the latitude $b$ being positive toward the north pole of the Moon.

The selenographic triad $\left\{{\bm A}, {\bm B}, {\bm C}\right\}$ is subject to a wobble and experiences small periodic variations in its spatial orientation called {\it physical libration} \citep{Viatteau2020}. The transformation between the triads $\left\{{\bm A}, {\bm B}, {\bm C}\right\}$ and $\left\{{\bm a}, {\bm b}, {\bm c}\right\}$ is given by 
\bal{bs3cp}
\left[{\bm A}{\bm B}{\bm C}\right]^{\rm T}&=&\left({\bm U}+\bm{\mathcal R}\right)\left[{\bm a}{\bm b}{\bm c}\right]^{\rm T}\;,
\ea
where the symbol T denotes vector transposition, ${\bm U}$ is the unit matrix, and $\bm{\mathcal R}$ is the matrix of the physical libration. Components of the libration matrix do not exceed 0.001~rad \citep{DE440_2021AJ} so that their effects in the calculation of function $\mathcal{F}_3(t)$ are negligibly small, and we can accept $\left\{{\bm A}, {\bm B}, {\bm C}\right\}=\left\{{\bm a}, {\bm b}, {\bm c}\right\}$.  
As mentioned above, the selenographic coordinates attached to the triad $\left\{{\bm A}, {\bm B}, {\bm C}\right\}$ rotate with respect to the BCRS making one revolution in one sidereal month around the rotational axis defined by the unit vector ${\bm c}$.  The velocity of the clock with respect to the LCRS is easily obtained by taking a time derivative of Eq. \eqref{sel8}, which yields
\bal{sel9}
\dot{\bm z}&=&n\rho\left({\bm B}\cos b\cos l+{\bm A}\cos b\sin l\right)\;,
\ea
where $n$ represents the mean motion of the Moon. When taking the derivative, we have neglected the radial oscillations and tangential movements of the Moon's surface as well as the precession of the ascending node.

Now we can calculate the function $\mathcal{F}_3$. It is achieved by making the dot product of the orbital velocity of the Moon, ${\bm v}=\dot{\bm r}$, with the selenographic radius-vector ${\bm z}$ of the lunar L-clock in definition \eqref{qq4}. The result is 
\bal{qq13}
\mathcal{F}_3(t)&=&\frac{na^2}{c^2}\frac{\rho}{a}\Bigl[\cos b\sin l-e\cos b\sin(M-l)-\sin(i+I)\sin b\cos F\Bigr]\;,
 \ea
where the angle $F:=M+\omega$.
Substituting the numerical values of the Earth-Moon system into the above equation yields, in nanoseconds (ns), 
\bal{zx5rj}
\mathcal{F}_3(t)&=&19.8\cos b\sin l-1.1\cos b\sin(M-l)-2.3\sin b\cos F\;\;{\rm ns}\;,
\ea
where for $\rho$ we have used 
the value of the mean radius of the Moon, $\rho=R_\L=1.737\times 10^3$ m. The first term in the right-hand side of formula \eqref{zx5rj} defines a constant shift in the readings of the lunar clocks having different selenographic coordinates, while the second and the third terms are periodic with a period of about 27.3 days. The periodicity is caused by the geometric libration of the Moon in longitude ($\sim e$) and latitude ($\sim\sin(i+I)$), respectively. The libration $\sin(M-l)$ signal in the function $\mathcal{F}_3(t)$ is maximal for clocks placed on the lunar equator, $b=0^\circ$, and vanishes for the clocks placed on the lunar poles, $b=\pm 90^\circ$. The $\cos F$ signal is  maximal for the clocks on the poles, $b=\pm 90^\circ$, and vanishes for the clocks on the lunar equator, $b=0^\circ$.

\section{Numerical Models}\label{s7}                      
\subsection{The TCL--TCG coordinate time transformation}\label{s7a}

In order to check the consistency of the overall theory and, in particular, the analytic formulas of Section~\ref{s5a}, we have numerically evaluated several time transformation formulas given in Sections~\ref{s1}, \ref{s2}, and \ref{s4}, using computer code that employs standard numerical integration techniques adapted specifically for this purpose.  We first tested the code
on the known TCB--TCG transformation formula \eqref{1} by comparing our results to an analytic approximation of this transformation given in the IAU Standards of Fundamental Astronomy (SOFA) \citep{sofa}, which is based on a development in Ref. \citep{fairhead90}.  (A more accurate analytical decomposition of the transformation is given in \citep{harada2003} but the SOFA code was adequate for our software validation.)  Differences over a 10-year integration were at the nanosecond level, which is similar to the estimated errors of the approximation.  Forward-backward tests on the integrations indicate numerical precision of $10^{-16}$.  The code utilizes the BCRS positions and velocities of the solar system bodies from the Jet Propulsion Laboratory's DE440 ephemeris \citep{DE440_2021AJ}.  Because that ephemeris uses Barycentric Dynamical Time (TDB) as its independent argument (with an average rate given approximately by the SI second on the geoid), rather than TCB, our numerical results have been divided by the scaling factor $(1 - L_\B)$, with $L_\B = 1.550519768 \times 10^{-8}$, as described in IAU resolution B1.5 of 2000 \citep{iau2000}. 

Figure~1 shows the results of the TCB--TCG integration over two years (out of ten years integrated).  A secular drift of 1.2794~ms/day has been removed\footnote{The computed secular drift is equivalent to the dimensionless constant $L_\C = 1-d({\rm TCG})/d({\rm TCB}) = 1.480 826 867\times10^{-8}$.}, leaving the prominent and well-known 1.6~ms annual term due to the Earth's orbital eccentricity.  The computer code allows for a choice of the solar system object that defines the body-centric reference system; the equivalent integration for the Moon, i.e., TCB--TCL as given by Eq. \eqref{2}, is shown in Fig.~2.  A secular drift of 1.2808~ms/day has been removed.  In Fig.~2, we see both the annual term seen in TCB--TCG and an additional monthly component, of order 0.1~ms, due to the Moon's orbit about the Earth.  In both of these computations, the location of the point in space where the time difference is computed is the center of mass of the body, i.e., the origin of the local reference system (GCRS or LCRS), so that the non-integrated location-dependent term in Eq. \eqref{1} or Eq. \eqref{2} is zero.  Both computations produce many smaller periodic components not obvious in the two figures.
\begin{figure}[tbp]
\centering
\includegraphics[scale=0.8]{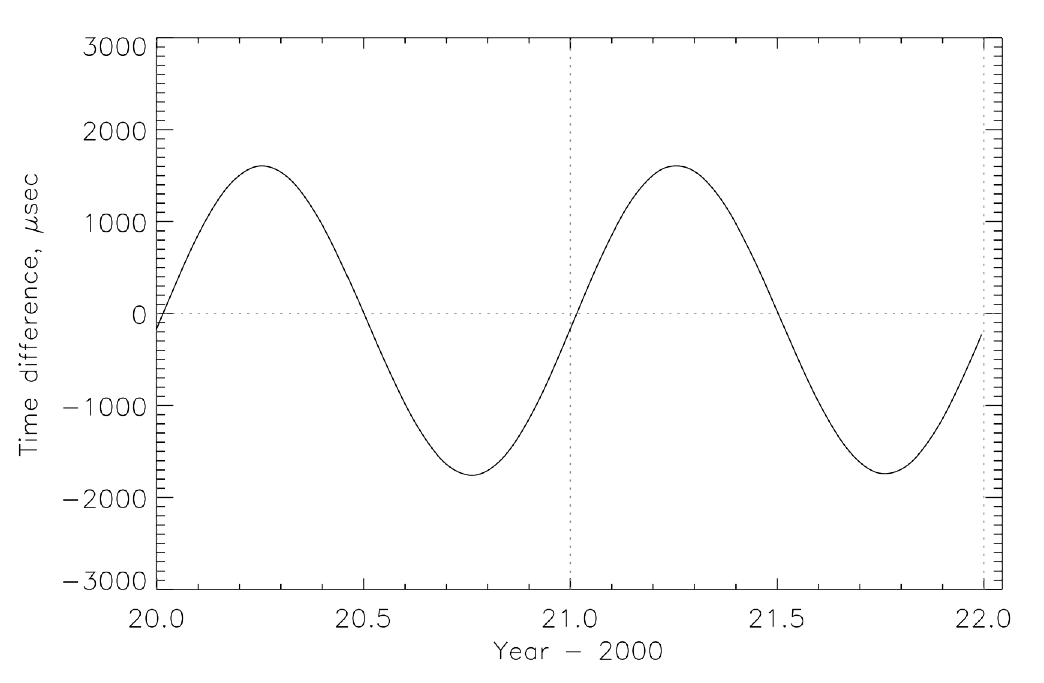}
\caption{The time difference TCB--TCG, computed at the geocenter.  The secular drift of 1.2794~ms/day has been removed.}
\label{fig1} 
\end{figure}
\begin{figure}[tbp]
\vspace{0.2in}
\includegraphics[scale=0.8]{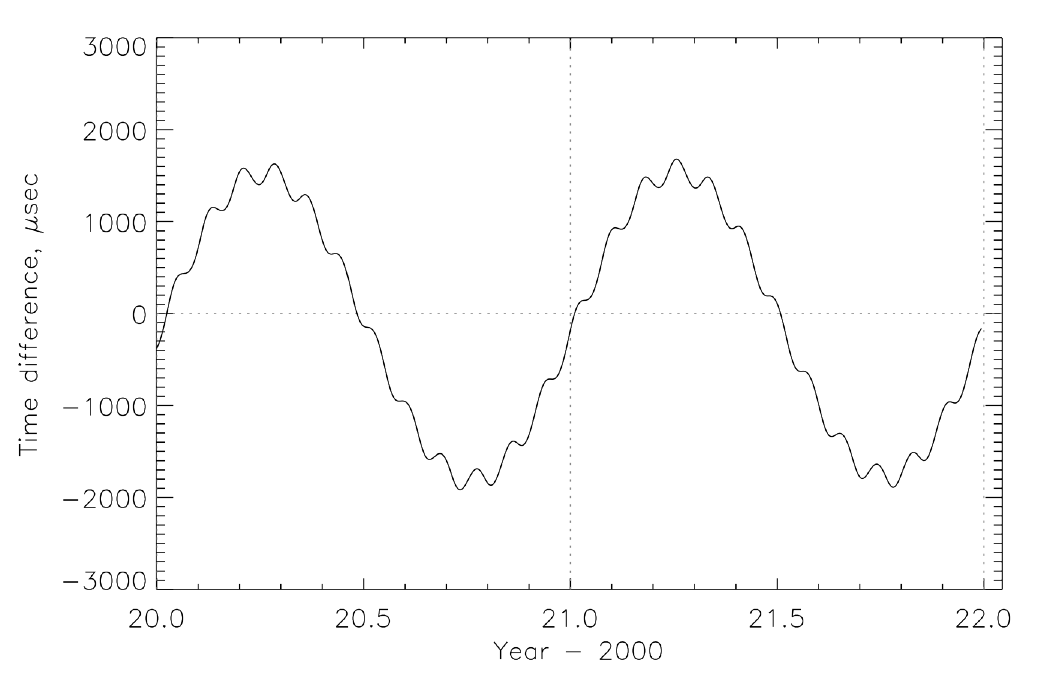}
\caption{The time difference TCB--TCL, computed at the center of mass of the Moon.  The secular drift of 1.2808~ms/day  has been removed.}
\label{fig2} 
\end{figure}

The result of the numerical integration of Eq.~\eqref{11} for TCL--TCG is shown in Fig. \ref{fig3}, where a linear rate term of $\approx\!-1.5\;\mu\rm{s/day}$ has been removed (more on the rate below).  For this integration, the point of reference is the center of mass of the Moon, so the last (non-integrated) term in Eq. \eqref{11} is zero.
\begin{figure}[tbp]
\centering
\includegraphics[scale=0.8]{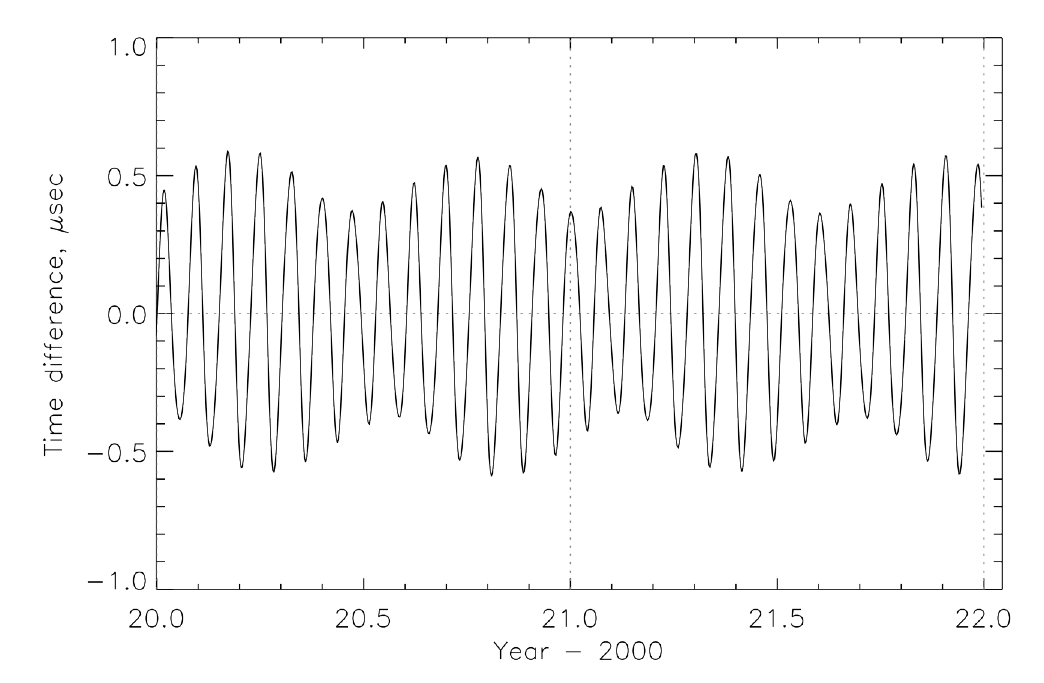}
\caption{Time difference TCL--TCG, with the secular drift removed, computed by evaluation of equation~\eqref{11} at the Moon's center of mass.} 
\label{fig3}
\end{figure}\noindent
The integration covered 10~years at 0.1-day intervals, although only the first two years are shown in Figure \ref{fig3}.   The integral in
Eq.~\eqref{11} for TCL--TCG is the same as that in Eq.~\eqref{a8} for LT--TT, so that all the resulting periodic components from the integration of 
\eqref{11} appear in both transformations.
The figure clearly shows the largest periodic term in the transformation shown in Eq. \eqref{num38} (for $\mathcal{F}_1 + \mathcal{F}_2$; $\mathcal{F}_3$ is zero in this case), which has an amplitude of $\approx\!0.5\;\mu\rm{s}$ and
a period of 27.55 days. The power spectrum of the periodicities is shown in Fig. \ref{fig4}, with the periods of the largest components listed in \eqref{pet52} and \eqref{num38} marked\footnote{The spectrum was generated using a conventional fast-Fourier transform (FFT) algorithm.  Because the FFT output samples the spectrum at discrete frequencies that may not exactly coincide with the frequencies present in the integration output, the relative heights of the peaks in the spectrum will not, in general, accurately relate to the actual differences in the amplitudes.}.
\begin{figure}[tbp]
\centering
\includegraphics[scale=0.8]{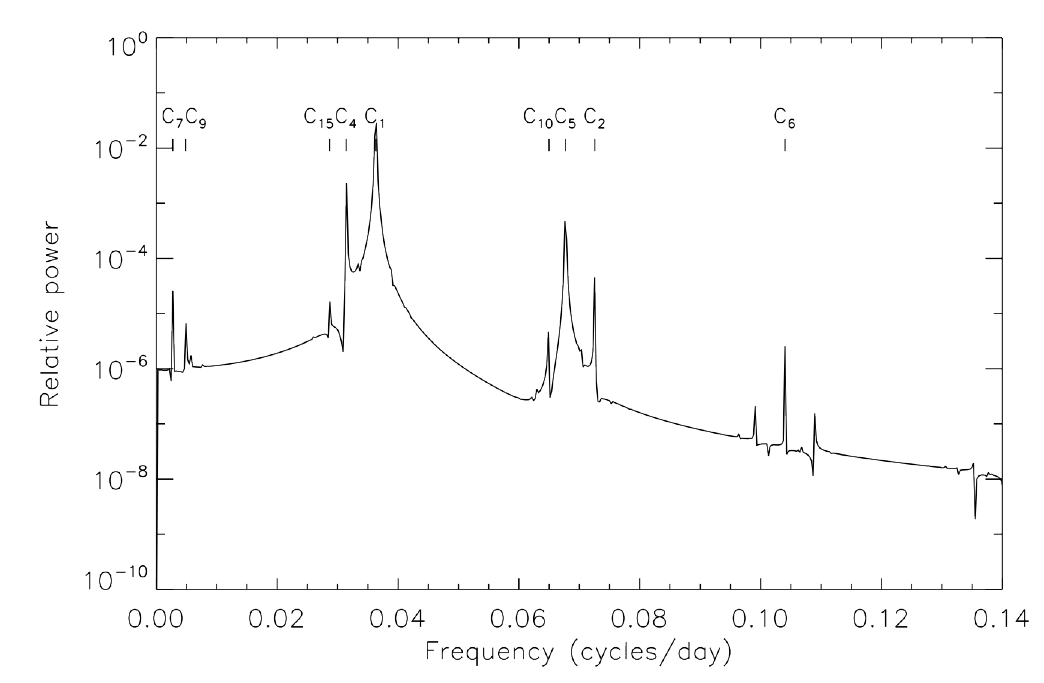}
\caption{The power spectrum of periodicities in the integration result shown in Fig. \ref{fig3}.}
\label{fig4} 
\end{figure}

Another numerical approach to  obtaining TCL--TCG is to subtract the results of the separate integrations of Eqs. \eqref{1} and \eqref{2}.  However, in doing so it is important that a common point of reference, ${\bm x}$, be used.  For ${\bm x}$ at the center of the Moon, i.e., ${\bm x} = {\bm x}_\L$, the last (non-integrated) term in Eq.~\eqref{1} must be evaluated but the corresponding term in Eq.~\eqref{2} is zero.  The overall difference between the coordinate times for the Moon and Earth time is then 
\bal{6v}
s-u&={\rm(TCB-TCG)}({\bm x}_L)-{\rm(TCB-TCL)}({\bm x_L}) = {\rm (TCL - TCG)}({\bm x_L}) \;,
\ea
The result is identical to the result from integrating Eq.~\eqref{11} except for a constant offset of $123.997\,\mu{\rm s}$, due to the different initializations in the two methods: all the integrations were arbitrarily set to zero at step~1 (on 2020 January~1), but Eq.~\eqref{1} yields a non-zero value at step~1 due to the the term outside the integral.
This independently validates Eq.~\eqref{11}.

If we extend the integration of Eq.~\eqref{11} to 30~years and solve for the amplitudes of the 15 periodic components listed in~\eqref{pet52}, and compare them to the analytically obtained amplitudes given in Eq.~\eqref{num38}, we obtain the results shown in Table~2.  
\begin{center}
\begin{tabular}{ccrcccc}
\multicolumn{7}{c}{\bf Table 2.  Major Periodic Components in TCL--TCG and LT--TT.} \\[0.08in] 
 Symbol  &        Luni-solar              & \w{xx}Period\w{i}  &\w{x}&    Analytic   & Numeric   &  Numeric    \\ 
(this paper) & \w{x}arguments\w{x}&                    && \w{x}amplitude\w{x} & \w{x}amplitude\w{x} & uncertainty \\[0.03in]
              &                                      & d\w{0000}     &&    $\mu$s    &   $\mu$s   &   $\mu$s     \\[0.01in] 
 $C_1$   &          $M$                      &  27.5546      &&   --0.4707   &  --0.4710    &    0.0003         \\        
 $C_2$   &        $2M$                      & 13.7773       &&   --0.0130   &  --0.0128    &    0.0001         \\       
 $C_3$   &        $3M$                      &    9.1848      &&   --0.0005   &  --0.0005    &  <0.0001\w{<} \\
 $C_4$  &         $2D-M$                  &  31.8119       &&  --0.0814   &  --0.0927    &    0.0002          \\
 $C_5$  &          $2D$                     &  14.7653       &&   --0.0468   & --0.0587    &    <0.0001\w{<} \\
 $C_6$  &        $2D+M$                  &    9.6137      &&  --0.0025    &  --0.0035    &    0.0001          \\
 $C_7$  &          $M'$                      & 365.2596     && \w{-} 0.0120 & \w{-}0.0100 &  0.0002        \\
 $C_8$  &        $2F-2D$                  &  173.3100  &&\w{-.}0.0005& \w{-}0.0013 &     0.0001         \\    
 $C_9$  &       $2D-2M$                  &  --205.8922  &&\w{-.}0.0009& --0.0046     &    0.0001          \\    
 $C_{10}$&         $2D-M'$               &  15.3873      &&  --0.0024    &  --0.0040    &    0.0001         \\    
 $C_{11}$&         $2D+M'$              &  14.1916      && --0.0003     &\w{-}0.0006  &     0.0001         \\    
 $C_{12}$&         $M-M'$                 &   29.8028    && --0.0016     &        ---         &       ---              \\    
 $C_{13}$&         $M+M'$                &   25.6217    &&\w{-.}0.0016 &\w{-}0.0023  &    0.0001         \\                   
 $C_{14}$&         $2D-M+M'$           &   29.2633   && \w{-} 0.0011 &      ---          &     ---             \\    
 $C_{15}$&         $2D-M-M'$            &   34.8469   &&  --0.0025      & --0.0041     &    0.0001                                    
\end{tabular}\\[0.05in]
\parbox{5in}{\small Here, $M$ is the Moon's mean anomaly, $M'$ is the Earth's mean anomaly,
$D$ is the mean elongation of the Moon from the Sun, and $F$ is the difference between the mean longitude of the Moon
and the longitude of the node of the lunar orbit ($F$ is also called the argument of latitude).  Negative amplitudes indicate that
the phase of the term is $\pm180^{\circ}$ from that indicated by the arguments.}
\end{center}

The formal uncertainties of the amplitudes from this 30-year solution are all <$0.0001\,\mu$s, and the phases were all very close to either 0 for positive amplitudes
or $\pm180^{\circ}$ for negative amplitudes.  The uncertainties listed for each component in the last column are half the total range of amplitude values ($\frac12$(max--min))
among five solutions involving 6-year subsets of the 30-year integration output, so they are more conservative than the formal errors.  We did not obtain reliable solutions for $C_{12}$ or $C_{14}$ (the phases were inconsistent among solutions).  The overall rate value computed in the same
30-year solution is $-1.4769\,\mu$s/day with an uncertainty of <$0.0001\,\mu$s/day, which can be compared to the analytical value
of $C_0 = -1.4714\,\mu$s/day given in Eq.~\eqref{bh4d}.
                                               
Due to the complexity of the lunar orbit, the table is undoubtedly incomplete.  Removing the rate term and these 15 components from the integration output
leaves a TCL--TCG time series that remains within a range of $\pm 7$\,ns and that shows evidence of other small periodicities.

\subsection{Clocks on the surfaces of the Earth and Moon}\label{s7c}

{The ultimate objective is to compare the proper time kept by hypothesized real clocks on the surfaces of the Moon and Earth.  The most appropriate comparison is between the coordinate times LT and TT, as given by Eqs. \eqref{a8}, \eqref{a8b}, and \eqref{a8a}, to which the proper time of any real clock on the surface of the Moon or Earth can be reduced.   If we limit the discussion to clocks on the reference equipotential surfaces, the geoid for the Earth and the selenoid for the Moon, then by Eqs.~\eqref{a8b} and \eqref{a8a}, the proper times reduce to TT for the Earth clock and LT for the Moon clock, neglecting the very small tidal potential terms.  It is important to note that there has been no international agreement on the definition of the selenoid (which may vary by lunar gravity model), so the precise rate of LT with respect to the coordinate time TCL may require future adjustment.}

The integrals in Eq.~\eqref{a8} have already been evaluated, since their sum is the same as the integral in Eq.~\eqref{11}, the computation of which was described in the previous subsection.  The results were described by a secular drift of  $-1.4769\,\mu$s/day and the periodic terms in Table~2.   Call that secular drift rate $L_{\sst{\rm{TCL-TCG}}}$; then the total rate difference between the Earth and Moon clocks is $L_\E - L_\L - L_{\sst{\rm{TCL-TCG}}} = 60.2147 - 2.7128 - 1.4769 = 56.025 \,\mu$s/day (TT day), using the rates for $L_\E$ and $L_\L$ given in Eqs. \eqref{11a} and \eqref{11b}.  The lunar clock runs faster than the terrestrial one by that amount, as a long-term average.

With the secular drift rate of proper time on the lunar surface with respect to the proper time on the Earth's geoid established, as well as the periodic terms given in Table~2 from the difference in the coordinate times at the center of mass of the Moon, the problem of comparing real clocks on the Earth and Moon reverts to the evaluation of the small
location-dependent component of TCL--TCG, or LT--TT, given by the last (non-integrated) term in Eq.~\eqref{a8}.  This component is approximated by the function $\mathcal{F}_3$ in the analytic development in Section \ref{F3} and involves a few nanosecond-level periodic components but no additional rate changes.
\begin{figure}[tbp]
\centering
\includegraphics[scale=0.8]{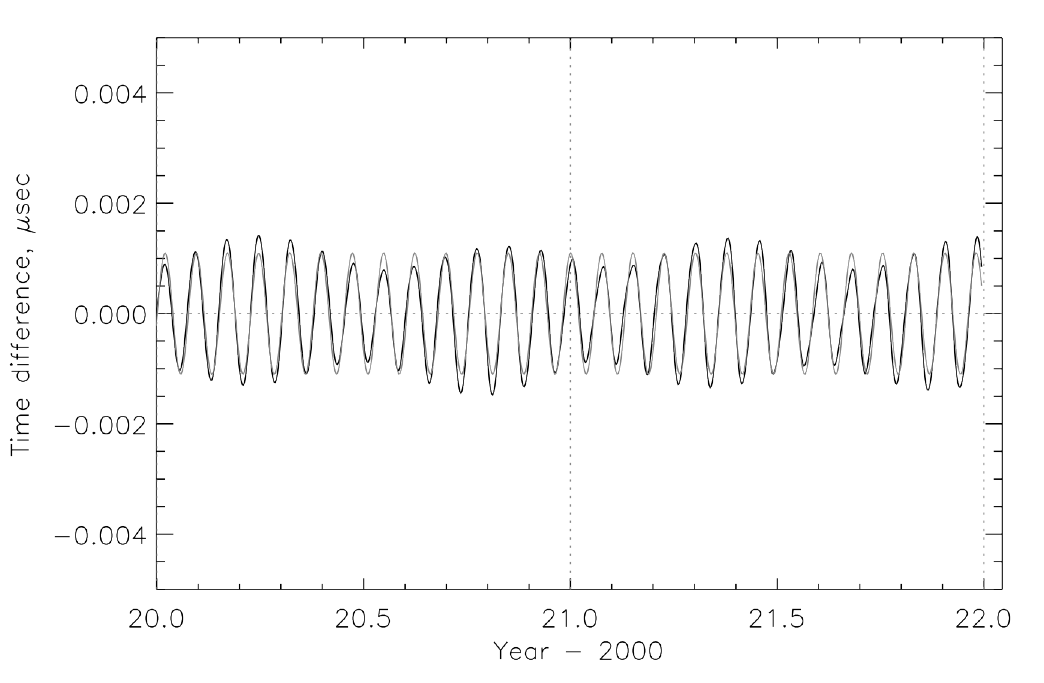}
\caption{Difference in time between TCL on the selenoid at a location near the center of the Moon's disk as seen from Earth (lunar longitude
$l=0^\circ$, latitude $b=0^\circ$) and TCL at the Moon's center of mass.  The location-dependent
term in \eqref{a8}, evaluated using the JPL ephemerides, is shown in black and the analytic approximation $\mathcal{F}_3$ is shown in gray.}
\label{fig5}
\end{figure}
\begin{figure}[tbp]
\centering
\includegraphics[scale=0.8]{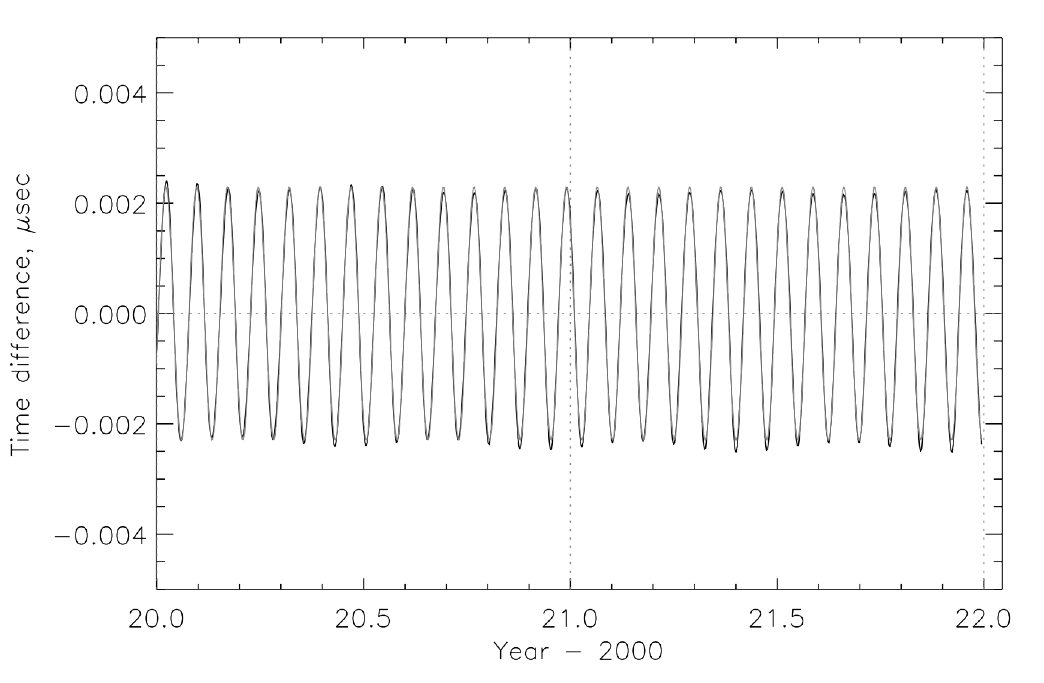}
\caption{Same as Figure \ref{fig5}, but for a clock at the lunar south pole. The two curves overlap almost exactly at this location.}
\label{fig6}
\end{figure}
\begin{figure}[tbp]
\centering
\includegraphics[scale=0.8]{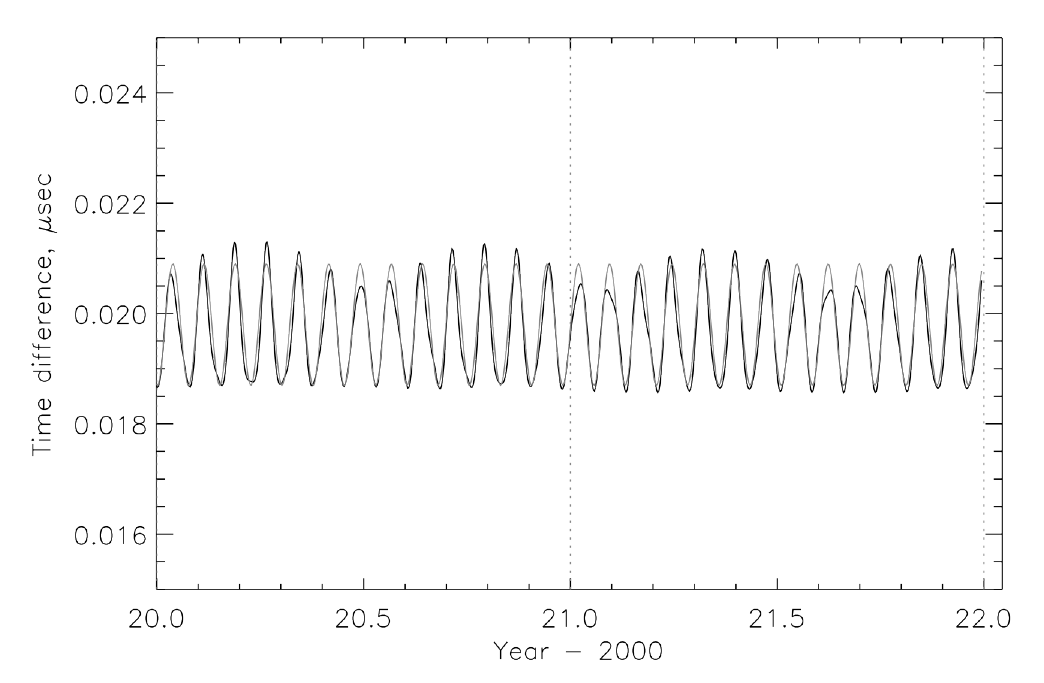}
\caption{ Same as Figure \ref{fig5}, but for a clock at the Moon's eastern limb (lunar longitude $l=90^\circ$, latitude $b=0^\circ$).}
\label{fig7}
\end{figure}
Evaluating the location-dependent term in Eq.~\eqref{a8}, using the lunar velocity and orientation data from the JPL DE440 ephemeris, we obtain Figs.~\ref{fig5}, \ref{fig6}, and \ref{fig7} for clocks at the center of the disk (longitude $l=0^\circ$, latitude $b=0^\circ$), the lunar south pole, and the eastern limb\footnote{``Eastern'' refers to an observer on the Moon; the eastern limb is toward the west in the Earth's sky.} at the equator (longitude $l=90^\circ$, latitude $b=0^\circ$), respectively.  These plots also show the curves representing the expression \eqref{zx5rj} for $\mathcal{F}_3$.  The center of disk location (Figure~\ref{fig5}) is sensitive to only the second term in \eqref{zx5rj}; the south pole location (Figure~\ref{fig6}) is sensitive to only the last term; and the equatorial limb location (Figure~\ref{fig7}) is sensitive to both the first and second terms.  As can be seen, the analytic expression well represents the numerical results.

\section{Conclusion}\label{s9}                            

The current decade will see increasing operations, both human and robotic, on the lunar surface and in orbit around the Moon.  A precise time standard,
or standards, for the Moon will be necessary to enable such basic capabilities as navigation, selenodesy, secure communications, and many types of scientific measurements \citep{gibney2023}.

We have examined the relativistic differences between hypothetical clocks on the Moon and Earth.   A Lunar Celestial Reference System has been described that is analogous to the Geocentric Celestial Reference System now used for precise applications on the Earth and in the near-Earth environment. 
TCL, a proposed lunar coordinate time for the LCRS, runs slow compared to TCG, geocentric coordinate time, by $1.477\,\mu\rm{s}/\rm{day}$ {at the origin of the LCRS} as a long-term average.
However, periodic variations at several amplitudes and frequencies, including a monthly component with an amplitude of $0.5\,\mu\rm{s}$, are also present.  LT, the {coordinate time on the adopted lunar selenoid, would run systematically fast compared to TT, the coordinate time on the adopted Earth geoid}, by $56.025\,\mu\rm{s}/\rm{day}$ as a long-term average.  The periodic variations that appear in the comparison of the two coordinate times are also present in the comparison of the two proper times, LT and TT, and may be significant for some practical applications. 

Satellite navigation systems require satellite clocks that are synchronized to $\approx$\,$30\,\rm{ns}$ for
$\approx$\,$10\,\rm{m}$ positioning accuracy, and that duplicate, at the same level of accuracy, a ground-based time standard that is used by the control system.\footnote{Satellite navigation systems such as GPS run on an internal time scale that has a known relation to UTC \citep{Ashby_2003LRR}.}   For such a
system for the Moon, this means that the satellite clocks must be synchronized within the LCRS, in TCL or a constant-rate-difference substitute for
TCL, i.e, $$t_{\rm nav} = {\rm TCL} \times (1 + r_{\rm nav}) + \Delta t_0\;,$$  where $r_{\rm nav}$ is the fractional rate offset ($|r_{\rm nav}| \ll 1$), and $\Delta t_0$ is an epoch difference.   That is, whatever time scale is used for navigation must advance by seconds that have a known and fixed relation to SI seconds {of TCL}.  Note, however, using any time scale for the LCRS that does not match the rate of TCL would require re-scaling the meter {to prevent unnecessary non-inertial forces from appearing in the equations of motion due to inconsistent units of time and length \citep{bk_1990,klioner2008}.} Furthermore, any astronomical constant present in equations where the meter is used as a dimension would also need to be recalibrated \citep{klioner2008,klioner2010}.

The differences between LT and TT described in this paper apply to any clock on the Moon that is set up to distribute Earth time.   Any clock system on the Moon that compensates in some way for the periodic variations described in Table~2
would produce seconds that would not have a constant length at the Moon.  For example, the $C_1$ periodic component represents a maximum
variation of 0.2\% in the LT--TT clock rate difference, and a maximum fractional variation in the length of a lunar second of $\sim\!10^{-12}$. Although this appears to be a small number, it is orders of mag\-nitude larger than the precision of the best modern atomic clocks.   This is an issue that may not be important for most ordinary lunar applications, but that would be a consideration for those in the future requiring very precise time or metrology.

\acknowledgments{We thank Valeri Makarov for helpful discussions and Neil Ashby for drawing our attention to his work on the lunar time \citep{Ashby_2024}. {We are  grateful to the anonymous referee for providing a highly constructive report that significantly helped clarify many essential theoretical points and improve the manuscript.} One of us (GHK) has been supported by U.S. Navy contract N0018921PZ202.} 

\bibliographystyle{unsrtnat}
\bibliography{Lunar_Time_Bibliography}

\begin{thebibliography}{56}
\providecommand{\natexlab}[1]{#1}
\providecommand{\url}[1]{\texttt{#1}}
\expandafter\ifx\csname urlstyle\endcsname\relax
  \providecommand{\doi}[1]{doi: #1}\else
  \providecommand{\doi}{doi: \begingroup \urlstyle{rm}\Url}\fi

\bibitem[{Israel} and {Esper}(2022)]{lunanet2022}
D.~J. {Israel} and J.~{Esper}.
\newblock \emph{LunaNet Interoperability Specification ({LNIS V4}).
  {ESC-LCRNS-SPEC-0015}}.
\newblock NASA Goddard Space Flight Center, Greenbelt, MD, December 2022.
\newblock URL
  \url{https://gs450drupal.gsfc.nasa.gov/static-files/LunaNet\%20Interoperability\%20Specification.pdf}.

\bibitem[Gibney(2023)]{gibney2023}
E.~Gibney.
\newblock What time is it on the {M}oon?
\newblock \emph{\nat}, 614:\penalty0 13--14, February 2023.
\newblock \doi{10.1038/d41586-023-00185-z}.

\bibitem[{Xie} and {Kopeikin}(2010)]{xiekop_2010}
Y.~{Xie} and S.~{Kopeikin}.
\newblock {Post-Newtonian Reference Frames for Advanced Theory of the Lunar
  Motion and a New Generation of Lunar Laser Ranging}.
\newblock \emph{\it Acta Physica Slovaca}, 60:\penalty0 393--495, August 2010.
\newblock \doi{10.2478/v10155-010-0004-0}.

\bibitem[{Kopeikin} and {Xie}(2010)]{kopeikin_2010CeMDA}
S.~{Kopeikin} and Y.~{Xie}.
\newblock {Celestial reference frames and the gauge freedom in the
  post-Newtonian mechanics of the Earth-Moon system}.
\newblock \emph{Celestial Mechanics and Dynamical Astronomy}, 108:\penalty0
  245--263, November 2010.
\newblock \doi{10.1007/s10569-010-9303-5}.

\bibitem[{Ashby} and {Patla}(2024)]{Ashby_2024}
N.~{Ashby} and B.~R. {Patla}.
\newblock {A Relativistic Framework to Estimate Clock Rates on the Moon}.
\newblock \emph{\aj}, 168\penalty0 (3):\penalty0 112, September 2024.
\newblock \doi{10.3847/1538-3881/ad643a}.

\bibitem[{Kopeikin} et~al.(2011){Kopeikin}, {Efroimsky}, and
  {Kaplan}]{kopeikin_2011book}
S.~{Kopeikin}, M.~{Efroimsky}, and G.~{Kaplan}.
\newblock \emph{Relativistic Celestial Mechanics of the Solar System}.
\newblock Wiley-VCH, Weinheim, September 2011.

\bibitem[{Rickman~ (Ed.)}(2001)]{iau2000}
H.~{Rickman~ (Ed.)}.
\newblock {Proceedings of the Twenty-Fourth General Assembly, Manchester, UK,
  2000}.
\newblock In \emph{Transactions of the International Astronomical Union},
  volume XXIV B, San-Francisco, USA, July 2001. {Astronomical Society of the
  Pacific}.
\newblock ISBN 1-58381-087-0.

\bibitem[{Kopeikin}(1988)]{Kopejkin_1988CeMec}
S.~M. {Kopeikin}.
\newblock {Celestial coordinate reference systems in curved space-time}.
\newblock \emph{Celestial Mechanics}, 44:\penalty0 87--115, March 1988.
\newblock \doi{10.1007/BF01230709}.

\bibitem[{Brumberg} and {Kopeikin}(1990)]{bk_1990}
V.~A. {Brumberg} and S.~M. {Kopeikin}.
\newblock {Relativistic time scales in the solar system}.
\newblock \emph{Celestial Mechanics and Dynamical Astronomy}, 48:\penalty0
  23--44, March 1990.
\newblock \doi{10.1007/BF00050674}.

\bibitem[{Soffel} et~al.(2003){Soffel}, {Klioner}, {Petit}, {Wolf}, {Kopeikin},
  {Bretagnon}, {Brumberg}, {Capitaine}, {Damour}, {Fukushima}, {Guinot},
  {Huang}, {Lindegren}, {Ma}, {Nordtvedt}, {Ries}, {Seidelmann},
  {Vokrouhlick{\'y}}, {Will}, and {Xu}]{soffel2003}
M.~{Soffel}, S.~A. {Klioner}, G.~{Petit}, P.~{Wolf}, S.~M. {Kopeikin},
  P.~{Bretagnon}, V.~A. {Brumberg}, N.~{Capitaine}, T.~{Damour},
  T.~{Fukushima}, B.~{Guinot}, T.-Y. {Huang}, L.~{Lindegren}, C.~{Ma},
  K.~{Nordtvedt}, J.~C. {Ries}, P.~K. {Seidelmann}, D.~{Vokrouhlick{\'y}},
  C.~M. {Will}, and C.~{Xu}.
\newblock {The IAU 2000 Resolutions for Astrometry, Celestial Mechanics, and
  Metrology in the Relativistic Framework: Explanatory Supplement}.
\newblock \emph{Astron. J.}, 126:\penalty0 2687--2706, December 2003.
\newblock \doi{10.1086/378162}.

\bibitem[{Soffel} and {Langhans}(2013)]{Soffel_2013book}
M.~{Soffel} and R.~{Langhans}.
\newblock \emph{{Space-Time Reference Systems}}.
\newblock Springer, Berlin, 2013.
\newblock \doi{10.1007/978-3-642-30226-8}.

\bibitem[nas()]{nasafacts}
{Planetary Fact Sheets}.
\newblock
  {\color{blue}\url{https://nssdc.gsfc.nasa.gov/planetary/planetfact.html}}.
\newblock Accessed: 2024-09-18.

\bibitem[{Soffel} and {Brumberg}(1991)]{sofbrum1991}
M.~H. {Soffel} and V.~A. {Brumberg}.
\newblock {Relativistic Reference Frames Including Timescales - Questions and
  Answers}.
\newblock \emph{Celestial Mechanics and Dynamical Astronomy}, 52\penalty0
  (4):\penalty0 355--373, December 1991.
\newblock \doi{10.1007/BF00048451}.

\bibitem[{Guinot}(1995)]{guinot1995}
B.~{Guinot}.
\newblock {Scales of Time}.
\newblock \emph{Metrologia}, 31\penalty0 (6):\penalty0 431--440, January 1995.
\newblock \doi{10.1088/0026-1394/31/6/002}.

\bibitem[{Capitaine} and {Guinot}(1997)]{capitaine1997}
N.~{Capitaine} and B.~{Guinot}.
\newblock {Reference systems in astronomy.}
\newblock \emph{Academie des Sciences Paris Comptes Rendus Serie B Sciences
  Physiques}, 11:\penalty0 725--738, June 1997.
\newblock \doi{10.1016/S1251-8069(97)83178-8}.

\bibitem[{Misner} et~al.(1973){Misner}, {Thorne}, and {Wheeler}]{mtw}
C.~W. {Misner}, K.~S. {Thorne}, and J.~A. {Wheeler}.
\newblock \emph{{Gravitation}}.
\newblock W.H.~Freeman, San Francisco, 1973.

\bibitem[{Kudryavtsev}(2016)]{Kudryavtsev-2016a}
S.~M. {Kudryavtsev}.
\newblock {Analytical series representing DE431 ephemerides of terrestrial
  planets}.
\newblock \emph{\mnras}, 456\penalty0 (4):\penalty0 4015--4019, March 2016.
\newblock \doi{10.1093/mnras/stv2892}.

\bibitem[{Kudryavtsev}(2017)]{Kudryavtsev-2016b}
S.~M. {Kudryavtsev}.
\newblock {Analytical series representing the DE431 ephemerides of the outer
  planets}.
\newblock \emph{\mnras}, 466\penalty0 (3):\penalty0 2675--2678, April 2017.
\newblock \doi{10.1093/mnras/stw3258}.

\bibitem[{Fienga} et~al.(2020){Fienga}, {Viswanathan}, {Deram}, {Di Ruscio},
  {Bernus}, {Laskar}, {Gastineau}, {Rambaux}, {Minazzoli}, {Durante}, and
  {Iess}]{INPOP_2020}
A.~{Fienga}, V.~{Viswanathan}, P.~{Deram}, A.~{Di Ruscio}, L.~{Bernus},
  J.~{Laskar}, M.~{Gastineau}, N.~{Rambaux}, O.~{Minazzoli}, D.~{Durante}, and
  L.~{Iess}.
\newblock {INPOP new release: INPOP19a}.
\newblock In C.~{Bizouard}, editor, \emph{Astrometry, Earth Rotation, and
  Reference Systems in the GAIA era}, pages 293--297, Paris, France, September
  2020. Observatoire de Paris.

\bibitem[{Park} et~al.(2021){Park}, {Folkner}, {Williams}, and
  {Boggs}]{DE440_2021AJ}
R.~S. {Park}, W.~M. {Folkner}, J.~G. {Williams}, and D.~H. {Boggs}.
\newblock {The JPL Planetary and Lunar Ephemerides DE440 and DE441}.
\newblock \emph{\aj}, 161\penalty0 (3):\penalty0 105, March 2021.
\newblock \doi{10.3847/1538-3881/abd414}.

\bibitem[{Fienga} and {Minazzoli}(2024)]{Fienga_2024LRR}
A.~{Fienga} and O.~{Minazzoli}.
\newblock {Testing theories of gravity with planetary ephemerides}.
\newblock \emph{Living Reviews in Relativity}, 27\penalty0 (1):\penalty0 1--99,
  January 2024.
\newblock \doi{10.1007/s41114-023-00047-0}.

\bibitem[{Charlot} et~al.(2020){Charlot}, {Jacobs}, {Gordon}, {Lambert}, {de
  Witt}, {B{\"o}hm}, {Fey}, {Heinkelmann}, {Skurikhina}, {Titov}, {Arias},
  {Bolotin}, {Bourda}, {Ma}, {Malkin}, {Nothnagel}, {Mayer}, {MacMillan},
  {Nilsson}, and {Gaume}]{icrf2020}
P.~{Charlot}, C.~S. {Jacobs}, D.~{Gordon}, S.~{Lambert}, A.~{de Witt},
  J.~{B{\"o}hm}, A.~L. {Fey}, R.~{Heinkelmann}, E.~{Skurikhina}, O.~{Titov},
  E.~F. {Arias}, S.~{Bolotin}, G.~{Bourda}, C.~{Ma}, Z.~{Malkin},
  A.~{Nothnagel}, D.~{Mayer}, D.~S. {MacMillan}, T.~{Nilsson}, and R.~{Gaume}.
\newblock {The third realization of the International Celestial Reference Frame
  by Very Long Baseline Interferometry}.
\newblock \emph{aap}, 644:\penalty0 A159, 2020.
\newblock \doi{10.1051/0004-6361/202038368}.
\newblock URL \url{https://doi.org/10.1051/0004-6361/202038368}.

\bibitem[{Malkin}(2024)]{Malkin_2024}
Z.~{Malkin}.
\newblock {How well is the International Celestial Reference System maintained
  in official IAU implementations?}
\newblock \emph{The Astronomical Journal}, 167\penalty0 (5):\penalty0 229,
  April 2024.
\newblock \doi{10.3847/1538-3881/ad35bf}.
\newblock URL \url{https://dx.doi.org/10.3847/1538-3881/ad35bf}.

\bibitem[{Kopeikin}(2019)]{k2019PRD}
S.~M. {Kopeikin}.
\newblock {Covariant equations of motion of extended bodies with arbitrary mass
  and spin multipoles}.
\newblock \emph{\prd}, 99\penalty0 (8):\penalty0 084008, April 2019.
\newblock \doi{10.1103/PhysRevD.99.084008}.

\bibitem[{Klioner} and {Voinov}(1993)]{Klioner_1993}
S.~A. {Klioner} and A.~V. {Voinov}.
\newblock {Relativistic theory of astronomical reference systems in closed
  form}.
\newblock \emph{\prd}, 48\penalty0 (4):\penalty0 1451--1461, August 1993.
\newblock \doi{10.1103/PhysRevD.48.1451}.

\bibitem[{Ashby}(2003)]{Ashby_2003LRR}
N.~{Ashby}.
\newblock {Relativity in the Global Positioning System}.
\newblock \emph{Living Reviews in Relativity}, 6\penalty0 (1):\penalty0 1--42,
  January 2003.
\newblock \doi{10.12942/lrr-2003-1}.

\bibitem[{Kouba}(2019)]{Kouba_2019GPSS}
J.~{Kouba}.
\newblock {Relativity effects of Galileo passive hydrogen maser satellite
  clocks}.
\newblock \emph{GPS Solutions}, 23:\penalty0 117, September 2019.
\newblock \doi{10.1007/s10291-019-0910-7}.

\bibitem[{Formichella} et~al.(2021){Formichella}, {Galleani}, {Signorile}, and
  {Sesia}]{Formichella_2021}
V.~{Formichella}, L.~{Galleani}, G.~{Signorile}, and I.~{Sesia}.
\newblock {Time-frequency analysis of the Galileo satellite clocks: looking for
  the $J_2$ relativistic effect and other periodic variations}.
\newblock \emph{GPS Solutions}, 25:\penalty0 56, February 2021.
\newblock \doi{10.1007/s10291-021-01094-2}.

\bibitem[{Wang} et~al.(2023){Wang}, {Li}, {Xue}, and {Xu}]{Wang_2023}
D.-X. {Wang}, M.~{Li}, H.-J. {Xue}, and T.-H. {Xu}.
\newblock {Analysis of the $J_2$ relativistic effect on the performance of
  on-board atomic clocks}.
\newblock \emph{GPS Solutions}, 27:\penalty0 114, April 2023.
\newblock \doi{10.1007/s10291-023-01453-1}.

\bibitem[{Kopeikin}(2007)]{kopeikin_2007PhRvL}
S.~M. {Kopeikin}.
\newblock Comment on ``{G}ravitomagnetic influence on gyroscopes and on the
  lunar orbit''.
\newblock \emph{Physical Review Letters}, 98\penalty0 (22):\penalty0 229001,
  June 2007.
\newblock \doi{10.1103/PhysRevLett.98.229001}.

\bibitem[{Brumberg} and {Kopejkin}(1989)]{brumkop1989NC}
V.~A. {Brumberg} and S.~M. {Kopejkin}.
\newblock {Relativistic reference systems and motion of test bodies in the
  vicinity of the Earth.}
\newblock \emph{Nuovo Cimento B}, 103\penalty0 (1):\penalty0 63--98, January
  1989.
\newblock \doi{10.1007/BF02888894}.

\bibitem[{Soffel} and {Han}(2019)]{soffel2019book}
M.~H. {Soffel} and W.-B. {Han}.
\newblock \emph{{Applied General Relativity}}.
\newblock Springer Nature, Switzerland, 2019.
\newblock \doi{10.1007/978-3-030-19673-8}.

\bibitem[iau()]{iau2024}
{Resolution of the XXXII IAU General Assembly to establish a standard Lunar
  Celestial Reference System (LCRS) and Lunar Coordinate Time (TCL)}.
\newblock
  {\color{blue}{\url{https://www.iau.org/static/resolutions/IAU2024_Resol2_English.pdf}}}.
\newblock Accessed: 2024-09-18.

\bibitem[{Ashby} and {Weiss}(2013)]{Ashby_2013}
N.~{Ashby} and M.~{Weiss}.
\newblock {Why there is no noon-midnight red shift in the GPS}.
\newblock \emph{arXiv e-prints}, art. arXiv:1307.6525, July 2013.
\newblock \doi{10.48550/arXiv.1307.6525}.

\bibitem[{Bothwell} et~al.(2019){Bothwell}, {Kedar}, {Oelker}, {Robinson},
  {Bromley}, {Tew}, {Ye}, and {Kennedy}]{Bothwell_2019}
T.~{Bothwell}, D.~{Kedar}, E.~{Oelker}, J.~M. {Robinson}, S.~L. {Bromley},
  W.~L. {Tew}, J.~{Ye}, and C.~J. {Kennedy}.
\newblock {JILA SrI optical lattice clock with uncertainty of $2.0 \times
  10^{-18}$}.
\newblock \emph{Metrologia}, 56\penalty0 (6):\penalty0 065004, oct 2019.
\newblock \doi{10.1088/1681-7575/ab4089}.
\newblock URL \url{https://dx.doi.org/10.1088/1681-7575/ab4089}.

\bibitem[Aeppli et~al.(2024)Aeppli, Kim, Warfield, Safronova, and
  Ye]{JILAclocks}
A.~Aeppli, K.~Kim, W.~Warfield, M.~S. Safronova, and J.~Ye.
\newblock {Clock with $8\times 10^{-19}$ Systematic Uncertainty}.
\newblock \emph{\prl}, 133:\penalty0 023401, Jul 2024.
\newblock \doi{10.1103/PhysRevLett.133.023401}.
\newblock URL \url{https://link.aps.org/doi/10.1103/PhysRevLett.133.023401}.

\bibitem[Agnew(2007)]{Agnew2007}
D.C. Agnew.
\newblock Earth tides.
\newblock In G.~Schubert, editor, \emph{Treatise on Geophysics}, pages 163 --
  195. Elsevier, Amsterdam, 2007.
\newblock ISBN 978-0-444-52748-6.
\newblock \doi{http://dx.doi.org/10.1016/B978-044452748-6.00056-0}.
\newblock URL
  \url{http://www.sciencedirect.com/science/article/pii/B9780444527486000560}.

\bibitem[{M{\"u}ller} et~al.(2018){M{\"u}ller}, {Dirkx}, {Kopeikin}, {Lion},
  {Panet}, {Petit}, and {Visser}]{Mueller_2018SSR}
J.~{M{\"u}ller}, D.~{Dirkx}, S.~M. {Kopeikin}, G.~{Lion}, I.~{Panet},
  G.~{Petit}, and P.~N.~A.~M. {Visser}.
\newblock {High Performance Clocks and Gravity Field Determination}.
\newblock \emph{\ssr}, 214\penalty0 (1):\penalty0 5, February 2018.
\newblock \doi{10.1007/s11214-017-0431-z}.

\bibitem[{Qin} et~al.(2020){Qin}, {Tan}, and {Shao}]{Qin_2020AJ}
C.-G. {Qin}, Y.-J. {Tan}, and C.-G. {Shao}.
\newblock {The Tidal Clock Effects of the Lunisolar Gravitational Field and the
  Earth's Tidal Deformation}.
\newblock \emph{\aj}, 160\penalty0 (6):\penalty0 272, December 2020.
\newblock \doi{10.3847/1538-3881/abc06f}.

\bibitem[{Bureau International des Poids et Mesures}(2019)]{BIPM2019}
{Bureau International des Poids et Mesures}.
\newblock \emph{{SI Brochure: The International System of Units (SI)}}.
\newblock BIPM, Paris, 9 edition, 2019.
\newblock URL
  \url{https://www.bipm.org/documents/20126/41483022/SI-Brochure-9-EN.pdf}.

\bibitem[{Fateev} et~al.(2015){Fateev}, {Kopeikin}, and {Pasynok}]{fateev_2015}
V.~F. {Fateev}, S.~M. {Kopeikin}, and S.~L. {Pasynok}.
\newblock {Effect of Irregularities in the Earth's Rotation on Relativistic
  Shifts in Frequency and Time of Earthbound Atomic Clocks}.
\newblock \emph{Measurement Techniques, Volume 58, Issue 6, p.~647-654},
  58:\penalty0 647--654, October 2015.
\newblock \doi{10.1007/s11018-015-0769-0}.

\bibitem[Torge and {M\"uller}(2012)]{torge2012}
W.~Torge and J.~{M\"uller}.
\newblock \emph{{Geodesy}}.
\newblock {De Gruyter}, {Berlin}, 2012.

\bibitem[{Hirt} and {Featherstone}(2012)]{Hirt_2012EPSL}
C.~{Hirt} and W.~E. {Featherstone}.
\newblock {A 1.5 km-resolution gravity field model of the Moon}.
\newblock \emph{Earth and Planetary Science Letters}, 329:\penalty0 22--30, May
  2012.
\newblock \doi{10.1016/j.epsl.2012.02.012}.

\bibitem[{Luzum} et~al.(2011){Luzum}, {Capitaine}, {Fienga}, {Folkner},
  {Fukushima}, {Hilton}, {Hohenkerk}, {Krasinsky}, {Petit}, {Pitjeva},
  {Soffel}, and {Wallace}]{Luzum11}
B.~{Luzum}, N.~{Capitaine}, A.~{Fienga}, W.~{Folkner}, T.~{Fukushima},
  J.~{Hilton}, C.~{Hohenkerk}, G.~{Krasinsky}, G.~{Petit}, E.~{Pitjeva},
  M.~{Soffel}, and P.~{Wallace}.
\newblock {The IAU 2009 system of astronomical constants: the report of the IAU
  working group on numerical standards for Fundamental Astronomy}.
\newblock \emph{Celestial Mechanics and Dynamical Astronomy}, 110\penalty0
  (4):\penalty0 293--304, August 2011.
\newblock \doi{10.1007/s10569-011-9352-4}.

\bibitem[{Ardalan} and {Karimi}(2014)]{Ardalan_2014CeMDA}
A.~A. {Ardalan} and R.~{Karimi}.
\newblock {Effect of topographic bias on geoid and reference ellipsoid of
  Venus, Mars, and the Moon}.
\newblock \emph{Celestial Mechanics and Dynamical Astronomy}, 118\penalty0
  (1):\penalty0 75--88, January 2014.
\newblock \doi{10.1007/s10569-013-9523-6}.

\bibitem[Kozai(1991)]{iau1992}
Y.~Kozai.
\newblock {Chapter II: Twenty-First General Assembly}.
\newblock \emph{Transactions of the International Astronomical Union},
  21\penalty0 (2):\penalty0 11--83, 1991.
\newblock \doi{10.1017/S0251107X00005459}.

\bibitem[{Kovalevsky}(1967)]{koval1967}
J.~{Kovalevsky}.
\newblock \emph{{Introduction to Celestial Mechanics}}.
\newblock {Astrophysics and Space Science Library, vol. 7, Reidel},
  {Dordrecht}, 1967.

\bibitem[{Brumberg}(1991)]{brum}
V.~A. {Brumberg}.
\newblock \emph{{Essential Relativistic Celestial Mechanics}}.
\newblock Adam Hilger, New York, 1991.

\bibitem[{Simon} et~al.(1994){Simon}, {Bretagnon}, {Chapront},
  {Chapront-Touze}, {Francou}, and {Laskar}]{Simon_1994}
J.~L. {Simon}, P.~{Bretagnon}, J.~{Chapront}, M.~{Chapront-Touze},
  G.~{Francou}, and J.~{Laskar}.
\newblock {Numerical expressions for precession formulae and mean elements for
  the Moon and the planets.}
\newblock \emph{\aap}, 282:\penalty0 663--683, February 1994.

\bibitem[{Murray}(1986)]{Murray_book}
C.~A. {Murray}.
\newblock \emph{{Vectorial astrometry.}}
\newblock {Adam Hilger LTD}, Bristol, 1986.

\bibitem[Viatteau(2020)]{Viatteau2020}
J.~Viatteau.
\newblock Lunar libration.
\newblock In B.~Cudnik, editor, \emph{Encyclopedia of Lunar Science}, pages
  1--3, Cham, 2020. Springer.
\newblock \doi{10.1007/978-3-319-05546-6_156-1}.

\bibitem[{IAU SOFA Board}(2021)]{sofa}
{IAU SOFA Board}.
\newblock {IAU SOFA Software Collection. Issue 2021-01-25}, 2021.
\newblock URL \url{http://www.iausofa.org}.

\bibitem[{Fairhead} and {Bretagnon}(1990)]{fairhead90}
L.~{Fairhead} and P.~{Bretagnon}.
\newblock {An analytical formula for the time transformation TB-TT.}
\newblock \emph{\aap}, 229:\penalty0 240--247, March 1990.

\bibitem[{Harada} and {Fukushima}(2003)]{harada2003}
W.~{Harada} and T.~{Fukushima}.
\newblock {Harmonic decomposition of time ephemeris TE405}.
\newblock \emph{\aj}, 126\penalty0 (5):\penalty0 2557--2561, November 2003.
\newblock \doi{10.1086/378909}.

\bibitem[{Klioner}(2008)]{klioner2008}
S.~A. {Klioner}.
\newblock {Relativistic scaling of astronomical quantities and the system of
  astronomical units}.
\newblock \emph{\aap}, 478:\penalty0 951--958, February 2008.
\newblock \doi{10.1051/0004-6361:20077786}.

\bibitem[{Klioner} et~al.(2010){Klioner}, {Capitaine}, {Folkner}, {Guinot},
  {Huang}, {Kopeikin}, {Pitjeva}, {Seidelmann}, and {Soffel}]{klioner2010}
S.~A. {Klioner}, N.~{Capitaine}, W.~M. {Folkner}, B.~{Guinot}, T.~Y. {Huang},
  S.~M. {Kopeikin}, E.~V. {Pitjeva}, P.~K. {Seidelmann}, and M.~H. {Soffel}.
\newblock {Units of relativistic time scales and associated quantities}.
\newblock In Sergei~A. {Klioner}, P.~Kenneth {Seidelmann}, and Michael~H.
  {Soffel}, editors, \emph{Relativity in Fundamental Astronomy: Dynamics,
  Reference Frames, and Data Analysis}, volume 261, pages 79--84, January 2010.
\newblock \doi{10.1017/S1743921309990184}.

\end{thebibliography}

\appendix
\section{The Earth-Moon Local Coordinate System and Lunar Time}\label{appendix}
In a recent scientific paper, \citet{Ashby_2024} propose a framework that addresses the establishment and dissemination of Coordinated Universal Time (UTC) from Earth to the Moon and beyond. Their approach utilizes a metric that is appropriate for a locally freely falling Earth-Moon system.  The approach by \citet{Ashby_2024} to the lunar time (LT) diverges from the formalism of local inertial coordinates, which is based on the IAU 2000 resolutions \citep{soffel2003} adopted in the present paper. Notably, it leads to the establishment of the LT--TT relationship \citep[Section 3]{Ashby_2024} that appears to differ formally from the result we obtained in Eq. \eqref{a8}. 

The primary objective of this appendix is threefold:
\begin{enumerate}
\item Extension of IAU 2000 resolutions on time scales: We aim to extend the IAU time resolutions to encompass the Earth-Moon system.
\item Comparison with the \citet{Ashby_2024} approach: We demonstrate that the LT--TT relationship derived through this extension not only reproduces but also generalizes the approach in \citet{Ashby_2024}.
\item Comparison with the standard approach based on the IAU 2000 resolutions: We demonstrate that the extended and standard approaches to establishing the coordinate time on the Moon lead to the identical LT--TT relationship and are interconvertible. It ensures consistency in timekeeping and facilitates seamless communication and navigation between Earth and the Moon.
\end{enumerate}

Throughout of the appendix we use some additional notations. Spatial vector and/or tensor indices $i,j,k$ take values 1,2,3. We use a multi-index tensor notation for {spatial Cartesian tensors and coordinates}, for example, a tensor $Q_K:=Q_{i_1i_2...i_k}$, a tensor $X^K:=X^{i_1i_2...i_k}=X^{i_1}X^{i_2}...X^{i_l}$. By the symbol $\pd_i$ we denote a partial derivative with respect to the $i$-th spatial coordinate. Multi-index notations are also used for high-order partial derivatives, for example, a partial derivative of the order $k$ is denoted
as $\pd_K:=\pd_{i_1i_2...i_k}$. A repeated spatial index represents a summation from~1 to~3, that is, $A_i B^i=A_1B^1+A_2B^2+A_3B^3$.

\subsection{Extension of IAU 2000 time resolutions to the Earth-Moon system}\label{appendix1}
The IAU 2000 resolutions concerning astronomical reference systems focus on two distinct coordinate systems within the solar system:  the Barycentric Celestial Reference System (BCRS) and the Geocentric Celestial Reference System (GCRS) and their respective coordinate times, TCB and TCG. The BCRS serves as a global coordinate chart that spans the entire spacetime. It provides a framework for precise astronomical measurements and calculations. Importantly, the BCRS metric takes into account the gravitational influence of all massive bodies in the solar system.
The GCRS, in contrast, represents an example of local inertial coordinates. It is constructed in the vicinity of a single massive body: our planet, Earth. Within this framework, the Moon is treated as one of the massive external bodies in the solar system. The Moon's orbital motion is considered on equal terms with Earth's orbital motion around the solar system barycenter.

However, the Moon is gravitationally bound to Earth and is an integral part of the Earth-Moon system. This system is in a state of free fall within the gravitational field of the Sun and other massive bodies in our solar system. In certain relativistic applications (such as establishing lunar time or conducting lunar laser ranging), where we need to treat the Moon as a massive body, it is more natural to extend the IAU reference system concepts to encompass the Earth-Moon system as a whole. To achieve this, we introduce an intermediate Earth-Moon coordinate chart, specifically designed for the Earth-Moon system and centered at the system's center of mass. 

Let us denote the Earth-Moon coordinate reference system (EMCRS) as $(T,{\bm X})$ where $T$ represents the new time coordinate and ${\bm X}=(X^i)=(X^1,X^2,X^3)$  are the spatial coordinates aligned with the coordinate axes of the BCRS. The EMCRS metric tensor accounts for the gravitational interaction between Earth and Moon. A detailed relativistic description of the EMCRS, including its metric tensor, is provided in the work by \citep{xiekop_2010}. This paper also establishes the connection between the EMCRS and other reference systems such as the BCRS, GCRS, and LCRS. By introducing the intermediate time scale $T$ associated with the Earth-Moon system, we can establish the lunar coordinate time (TCL) and explore the TCL--TCG time transformation from a different perspective. 

Firstly, let's establish the transformation between the time $T$ of the Earth-Moon system and $t=$TCB (Barycentric Coordinate Time).  The derivation of this transformation relies upon the same guiding principles of the IAU resolutions as the derivation of the transformation between TCG (Geocentric Coordinate Time) and TCB. However, in this case, we replace the Earth with the Earth-Moon system. The relevant equation is given as follows \citep[Eq. 5.24]{xiekop_2010} (notice that designations used for time and space coordinates in paper \citep{xiekop_2010} differ from the present paper):
\bal{ap1}
T&=&t-\frac{1}{c^2}\int\limits_{t_0}^t\left[\frac{v_\B^2}{c^2}+\xoverline{U}\left({\bm x}_\B\right)\right]dt-\frac{1}{c^2}{\bm v}_\B\cdot{\bm r}_\B\;,
\ea
where
\bal{ap2}
\xoverline{U}\left({\bm x}\right)&=&\sum_{\A\neq\E,\L}\frac{\mu_\A}{r_{\A}}\;,
\ea
is the gravitational potential of the external bodies taken at the point ${\bm x}={\bm x}_\B$, $r_A=|{\bm r}_\A|$, and ${\bm r}_\A={\bm x}-{\bm x}_\A$. Notice that both Earth and Moon have been excluded from the potential $\xoverline{U}$. Transformation of the spatial coordinates of the BCRS and EMCRS is given by a simple translation of the coordinate origin
\bal{et5a}
{\bm X}={\bm x}-{\bm x}_\B\;,
\ea
where ${\bm x}_\B$ are the BCRS coordinates of the center of mass of the Earth-Moon system. In what follows, two EMCRS vectors will be important: (1) the radius vector of the Earth's center of mass, ${\bm X}_\E={\bm x}_\E-{\bm x}_\B$, and (2) the radius vector of the Moon's center of mass, ${\bm X}_\L={\bm x}_\L-{\bm x}_\B$. It is worth noting that the EMCRS spatial vector ${\bm X}_{\L\E}$ is equivalent to the geocentric radius-vector of the Moon ${\bm r}_{\L\E}$, that is ${\bm X}_{\L\E}={\bm r}_{\L\E}$ which can be checked by inspection.

Both the GCRS and LCRS exists in a state of free fall with respect to the barycenter of the Earth-Moon system. Since the Earth and Moon are gravitationally bound, the rate of the E-clock on the Earth is subject to the direct influence of the gravitational potential of the Moon. However, because the Earth-Moon system is itself in a state of free fall with respect to the barycenter of the solar system, the gravitational field of other bodies of the solar system can appear only in the form of a tidal potential. Hence, the time transformation between $u=$TCG (Geocentric Coordinate Time) and the coordinate time $T$ of the Earth-Moon coordinate system takes on the following form \citep[Eq. 5.63]{xiekop_2010}: 
\bal{ap3}
u&=&T-\frac{1}{c^2}\int\limits_{T_0}^T\left[\frac{V_\E^2}{2}+\frac{\mu_\L}{R_{\L\E}}+Q_iX^i_\E+\sum_{k=2}^\infty\frac{1}{k!}Q_K X^K_\E\right]dT-\frac{1}{c^2}{\bm V}_\E\cdot{\bm R}_\E\;,
\ea
where ${\bm V}_\E=\dot{\bm X}_\E$ is the velocity of the geocenter with respect to the barycenter of the Earth-Moon system, ${\bm X}_\E=(X^i_\E)$ represents the EMCRS coordinate of the geocenter, ${\bm R}_\E={\bm X}-{\bm X}_\E={\bm r}_\E$ is the radius-vector from the geocenter to the point of measurement of TCG, and ${\bm R}_{\L\E}={\bm X}_\L-{\bm X}_\E={\bm r}_{\L\E}$ is the radius-vector from geocenter to the center of mass of the Moon. 

Equation \eqref{ap3} contains additional parameters compared to the transformation between TCG and TCB. They include the dipole moment, $Q_i$, and higher-order external multipole moments, $Q_K:=Q_{i_1i_2...i_k}$ ($k\le 2$). The dipole moment $Q_i$ is interconnected to the high-order multipoles by the constraint that the linear momentum of the Earth-Moon system vanishes in the local EMCRS coordinates \citep[Section 6.1.4]{kopeikin_2011book}. We obtain
\bal{ap4}
Q_i&=&\-\left(M_\E+M_\L\right)^{-1}\sum_{k=1}^\infty Q_{ii_1i_2...i_k}M^{i_1i_2...i_k}\;,
\ea
where $M^{i_1i_2...i_k}$ are the intrinsic multipole moments of the gravitational field of the Earth-Moon system. {In the monopole-mass approximation these multipole moments can be explicitly expressed as a linear combination of two terms. These terms depend solely on the masses of the Earth and Moon, as well as the coordinates of their centers of mass,}  
\bal{dipo32}
M^{i_1i_2...i_k}&=&M_\E X_\E^{i_1i_2...i_k}+M_\L X_\L^{i_1i_2...i_k} \qquad\quad(k\ge 2) \;.
\ea
{The residual terms omitted from Eq. \eqref{dipo32} depend on the internal multipole moments of the Earth and Moon. Their contribution to each term on the right-hand side of Eq. \eqref{dipo32} is reduced by a factor corresponding to the oblateness coefficient $J_2$ of the Earth and Moon, which are approximately $1.08\times 10^{-3}$ and $0.2\times 10^{-3}$, respectively -- see \citep{nasafacts}.}

Thus, physical significance of the dipole moment $Q_i$ is that
it describes how the combined center of mass of the Earth and Moon (the Earth-Moon barycenter) accelerates relative to a hypothetical particle which is moving freely along a path (called a timelike geodesic) in the gravitational field created by the Sun and other large bodies in the solar system. The acceleration is caused by the gravitational coupling of the intrinsic multipole moments of the Earth-Moon system
$M^{i_1i_2...i_k}$
with the external multipole moments of tidal gravitational field of the external bodies of the solar system, which are defined in terms of the partial derivatives
\ba\label{ap5}
Q_{i_1i_2...i_k}&=&\pd_{i_1i_2...i_k}{\xoverline U}({\bm x}_\B)\;,
\ea
of the external gravitational field of the solar system bodies computed at the barycenter, ${\bm x}_\B$, of the Earth-Moon system.  The motion of the center of mass of the Earth-Moon system with respect to BCRS obeys the equations of motion 
\bal{ghy76}
a^i_\B&=&\pd_i{\xoverline U}({\bm x}_\B)-Q_i\;.
\ea
where $(a^i_\B)={\bm a}_\B=\ddot{\bm x}_\B$ is the BCRS coordinate acceleration of the Earth-Moon barycenter. 

Similar arguments are applied to the derivation of the time transformation between TCL$= s$ and the coordinate time $T$ of the Earth-Moon system.  We have \citep[Eq. 5.78]{xiekop_2010}
\bal{ap6}
s&=&T-\frac{1}{c^2}\int\limits_{T_0}^T\left[\frac{V_\L^2}{2}+\frac{\mu_\E}{R_{\L\E}}+Q_iX^i_\L+\sum_{k=2}^\infty\frac{1}{k!}Q_K X^K_\L\right]dT-\frac{1}{c^2}{\bm V}_\L\cdot{\bm R}_\L\;,
\ea
where ${\bm X}_\L=(X^i_\L)$ is the EMCRS coordinate of the Moon's center of mass,  ${\bm V}_\L=\dot{\bm X}_\L$ is the velocity of the Moon with respect to the barycenter of the Earth-Moon system, ${\bm R}_\L={\bm X}-{\bm X}_\L={\bm r}_\L$ is the radius-vector from the center of mass of the Moon to the point of measurement of TCL, and the dipole moment $Q_i$ is given in Eq.~\eqref{ap4} and the multipole moments $Q_K$ in Eq.~\eqref{ap5}.

The infinite sums appearing in Eqs. \eqref{ap3} and \eqref{ap6} can be considered part of the Taylor expansion of the gravitational potential $\xoverline U({\bm x})$ of the external bodies with the center of the expansion taken at different points of space. Indeed, it is straightforward to confirm that
\bal{ap7}
\sum_{k=2}^\infty\frac{1}{k!}Q_K X^K_\E&=&{\xoverline U}({\bm x}_\E)-\xoverline U({\bm x}_\B)-X^i_\E\pd_i\xoverline U({\bm x}_\B)\;,
\\\label{ap8}
\sum_{k=2}^\infty\frac{1}{k!}Q_K X^K_\L&=&{\xoverline U}({\bm x}_\L)-\xoverline U({\bm x}_\B)-X^i_\L\pd_i\xoverline U({\bm x}_\B)\;.
\ea

Subtracting \eqref{ap3} from \eqref{ap6} cancels out the coordinate time $T$ and leads to a new equation that describes the relationship between TCL (Lunar Coordinate Time) and TCG (Geocentric Coordinate Time)
\ba\nonumber
s&=&u-\frac{1}{c^2}\int\limits_{T_0}^T\left[\frac{V_\L^2-V_\E^2}{2}+\frac{\mu_\E-\mu_\L}{r_{\L\E}}+\xoverline U({\bm x}_\L)-\xoverline U({\bm x}_\E)+r_{\L\E}^iQ_i-r_{\L\E}^i\pd_i\xoverline U({\bm x}_\B)\right]dT\\\label{ap9}&&~~
-\frac{1}{c^2}\left({\bm V}_\L\cdot{\bm r}_\L-{\bm V}_\E\cdot{\bm r}_\E\right),
\ea
where $(r_{\L\E}^i)={\bm r}_{\L\E}={\bm X}_{\L\E}$ is the radius-vector of the Moon with respect to the Earth's geocenter.
This equation can be elaborated further by expanding the gravitational potential $\xoverline U({\bm x}_\L)$ in a Taylor series around the point ${\bm x}_\E$. This yields
\ba\nonumber
s&=&u-\frac{1}{c^2}\int\limits_{T_0}^T\left[\frac{V_\L^2-V_\E^2}{2}+\frac{\mu_\E-\mu_\L}{r_{\L\E}}+W({\bm x}_\E)+r_{\L\E}^iQ_i+r_{\L\E}^i\pd_i\xoverline U({\bm x}_\E)-r_{\L\E}^i\pd_i\xoverline U({\bm x}_\B)\right]dT
\\\label{ap10}
&&~~-\frac{1}{c^2}\Bigl({\bm V}_\L\cdot{\bm r}_\L-{\bm V}_\E\cdot{\bm r}_\E\Bigr)\;,
\ea  
where the perturbing tidal potential of the external bodies
\bal{ap11}
W({\bm x}_\E)&:=&\sum_{k=2}^\infty\frac{1}{k!}\pd_{i_1i_2...i_k}\xoverline U({\bm x}_\E)r_{\L\E}^{i_1i_2...i_k}\;,
\ea
is computed at the geocenter.
Finally, we expand the difference between the two last terms in the integrand of Eq. \eqref{ap10} in the Taylor series
\bal{ap12}
r_{\L\E}^i\pd_i\xoverline U({\bm x}_\E)-r_{\L\E}^i\pd_i\xoverline U({\bm x}_\B)&=&\sum_{k=1}^\infty\frac{1}{k!}Q_{iK}X^K_\E\;,
\ea
and express the coordinate times $s=$TCL and $u=$TCG in terms of proper times LT and TT measured on selenoid and geoid, respectively, by means of
Eqs.~\eqref{a91a}, \eqref{a92b}. It allows us to present Eq.~\eqref{ap10} in the following form:
\ba\nonumber
{\rm LT}&=&\left(1+L_\E-L_\L\right){\rm TT}-\frac{1}{c^2}\int\limits_{t_0}^t\left[\frac12\left(V_\L^2-V_\E^2\right)+\frac{\mu_\E-\mu_\L}{r_{\L\E}}+W({\bm x}_\E)+r_{\L\E}^i\left(Q_i+\sum_{k=1}^\infty\frac{1}{k!}Q_{iK}X^K_\E\right)\right]dt
\\\label{ap13}
&&\phantom{\left(1+L_\E-L_\L\right){\rm TT}}-\frac{1}{c^2}\Bigl({\bm V}_\L\cdot{\bm r}_\L-{\bm V}_\E\cdot{\bm r}_\E\Bigr)\;,
\ea 
where we have replaced integration with respect to time $T$ with integration with respect to TCB time $t$, because the difference between these two time scales is of the post-Newtonian order of magnitude and can be neglected.

Equation \eqref{ap13} is fundamental in the extension of the IAU time resolutions to the Earth-Moon system. It expresses the lunar time in terms of the terrestrial time which is directly connected to UTC, and all quantities entering the LT--TT transformation are measured by a fictitious observer who is at rest with respect to the Earth-Moon barycenter. 

\subsection{Comparison with the Ashby and Patla approach}\label{appendix2}

The fundamental time transformation Eq. \eqref{ap13} can be compared to the corresponding equation derived by \citet{Ashby_2024}. These authors considered the fractional frequency shift of two atomic clocks counting the lunar and terrestrial time, but they neglected all tidal terms that appear in Eq.~\eqref{ap13} under the sign of the integral. The proper time of the L-clock on the Moon, $\tau_\L$, and the proper time of the E-clock on Earth, $\tau_\E$, are linear functions of the lunar time LT and terrestrial time TT,
respectively, as evident from Eqs.~\eqref{a8b}, \eqref{a8a}. Making use of these equations {to convert LT and TT to the proper times $\tau_\L$ and $\tau_\E$ respectively, and differentiating Eq.~\eqref{ap13}, with all tidal, height-dependent, and multipole-moment terms discarded}, we obtain
\bal{ap14}
\frac{d\tau_\L-d\tau_\E}{d\tau_\E}&=&-\frac{1}{2c^2}\Bigl(V^2_\L-V^2_\E\Bigr)+\frac{\mu_\L-\mu_\E}{c^2r_{\L\E}}+\frac{\Phi_{0\E}-\Phi_{0\L}}{c^2}\\\nonumber
&&-\frac{1}{c^2}\Bigl({\bm A}_\L\cdot{\bm z}-{\bm A}_\E\cdot{\bm w} \Bigr)-\frac{1}{c^2}\Bigl({\bm V}_\L\cdot\dot{\bm z}-{\bm V}_\E\cdot\dot{\bm w} \Bigr)\;,
\ea
where the notation for the selenocentric (LCRS) coordinates ${\bm z}={\bm r}_\L$ and that for the geocentric (GCRS) coordinates ${\bm w}={\bm r}_\E$ of the atomic clocks have been used, and ${\bm A}_\E=\dot{\bm V}_\E$ and ${\bm A}_\L=\dot{\bm V}_\L$ are the coordinate accelerations of the Earth and Moon with respect to the barycenter of the Earth-Moon system.
Terms in the first line of the right-hand side of Eq. \eqref{ap14} coincide with the terms given by \citet{Ashby_2024} in their equation (18), while the terms from the second line of Eq. \eqref{ap14} appear in equations (29), (32) and (33) of their paper. This highlights a valuable correspondence between the paper by \citet{Ashby_2024} and the extended IAU 2000 resolutions on time scales described in this appendix. The extended IAU 2000 framework provides a more fundamental approach to the time scales in the Earth-Moon system by accounting not only for the kinematic (motion-related) effects and those from the gravitational field of the Earth and Moon, but it also includes a comprehensive description of the tidal effects on the time transformations, which helps us to better understand how we measure time.  

\subsection{Comparison with the standard approach based on the IAU 2000 resolutions}\label{appendix3}

Let us compare Eq. \eqref{ap13} with Eq. \eqref{a8}, which has been obtained in the main body of the present paper without introducing the intermediate local coordinates attached to the barycenter of the Earth-Moon system. The kinematic (velocity-dependent) term under the sign of integral in Eq.~\eqref{ap13} can be decomposed into three
terms:
\bal{ap15}
\frac{V_\L^2-V_\E^2}{2}&=&\frac{v_{\L\E}^2}{2}-{\bm A}_\E\cdot{\bm r}_{\L\E}+\frac{d}{dT}\Bigl({\bm V}_\E\cdot{\bm r}_{\L\E}\Bigr)\;,
\ea
where ${\bm r}_{\L\E}:={\bm X}_\L-{\bm X}_\E={\bm x}_\L-{\bm x}_\E$ is the radius-vector of the Moon with respect to Earth's geocenter, ${\bm v}_{\L\E}:={\bm V}_\L-{\bm V}_\E={\bm v}_\L-{\bm v}_\E$ is the relative velocity of the Moon with respect to Earth, and ${\bm A}_\E=\dot{\bm V}_\E$ is the coordinate acceleration of the geocenter with respect to the barycenter of the Earth-Moon system. Interestingly, the acceleration ${\bm A}_\E$ of the geocenter obeys the Newtonian equation of motion valid in the EMCRS coordinates (c.f. \citep[Eq. 5.73]{xiekop_2010})
\bal{ap16}
A^i_\E&=&\frac{\mu_\L}{r_{\L\E}^3}r_{\L\E}^i+Q_i+\sum_{k=1}\frac{1}{k!}Q_{iK}X^K_\E\;.
\ea
Replacing Eqs. \eqref{ap15}, \eqref{ap16} in Eq. \eqref{ap13}, integrating the term with the total time derivative, and using the identity 
\bal{ap17}
{\bm V}_\E\cdot{\bm r}_{\L\E}+{\bm V}_\L\cdot{\bm r}_\L-{\bm V}_\E\cdot{\bm r}_\E&=&{\bm v}_{\L\E}\cdot{\bm z}\;,
\ea
{where ${\bm z}={\bm x}-{\bm x}_\L$ is a selenocentric position of the L-clock}, results in
\bal{ap18}
{\rm LT}&=&\Bigl(1+L_\E-L_\L\Bigr){\rm TT}-\frac{1}{c^2}\int\limits_{t_0}^t\left(\frac{v_{\L\E}^2}{2}+\frac{{\mu}_\E-2\mu_\L}{r_{\L\E}}+W\right)dt-\frac{1}{c^2}{\bm v}_{\L\E}\cdot{\bm z}\;,
\ea
which exactly coincides with Eq. \eqref{a8} derived without introducing the intermediate Earth-Moon local coordinates. 

Taking the time derivative of Eq. \eqref{ap18}, we obtain both a significant simplification and a generalization of \citet{Ashby_2024} formula, c.f. Eq. \eqref{ap14}
\bal{htrc5}
\frac{d\tau_\L-d\tau_\E}{d\tau_\E}&=&-\frac{v_{\L\E}^2}{2c^2}+\frac{\mu_\L-2\mu_\E}{c^2r_{\L\E}}+\frac{\Phi_{0\E}-\Phi_{0\L}}{c^2}+\frac{W}{c^2}-\frac{1}{c^2}\Bigl(\dot{\bm v}_{\L\E}\cdot{\bm z}-{\bm v}_{\L\E}\cdot\dot{\bm z} \Bigr)\;.
\ea 
Notice that the tidal potential $W$ entering Eqs. \eqref{ap18}, \eqref{htrc5} has been defined in Eq. \eqref{ap11}. It generalizes the tidal potential $W$ from Eq. \eqref{a8} by accounting for all other gravitating bodies of the solar system (besides the Sun) lying external to the Earth-Moon system.

The derivation of the TCL--TCG relationship presented in this appendix validates the equivalence between two methodologies. The first method directly employs the IAU 2000 resolutions framework, while the second incorporates the intermediate freely-falling local coordinates of the Earth-Moon system.

Formula \eqref{a8} offers a distinct advantage by working with directly measurable orbital parameters of the lunar orbit, such as the relative Earth-Moon distance, ${\bm r}_{\L\E}$, and the geocentric velocity of the Moon, ${\bm v}_{\L\E}$.

In contrast, Eq. \eqref{ap13} requires knowledge of the not directly measurable velocities of Earth, ${\bm V}_\E$, and Moon, ${\bm V}_\L$, with respect to the barycenter of the Earth-Moon system.  Additionally, Eq.~\eqref{ap13} includes a more complex description of the tidal terms compared to Eq.~\eqref{a8} or Eq.~\eqref{ap18}. 
\end{document}